\documentclass[seceq]{ptptex}

\usepackage{graphicx}

\title{\bf Correlated EoM and Distributions for A=6 Nuclei}

\author{M. Tomaselli$^{1,2}$, T. K{\"u}hl$^2$, D. Ursescu$^2$, and
S. Fritzsche$^3$}

\inst{$^1$ Institute of Physics, Darmstadt University, D-64289 Darmstadt, Germany \\ 
$^2$ GSI Gesellschaft f{\"u}r Schwerionenforschung, D-64291 Darmstadt,Germany\\
$^3$ Institute of Physics, Kassel University, D-34132 Kassel, Germany}

\abst{
Energy spectra and electromagnetic transitions of nuclei are strongly depending
from the correlations of the bound nucleons. Two particle correlations are responsible for
the scattering of model particles either to low momentum-
or to high momentum-states. 
The low momentum states form the model space while the high momentum states
are used to calculate the G-matrix. 
The three and higher order particle correlations do not play a role in
the latter calculation especially if the correlations induced
by the scattering operator are of sufficient short range. They modify however, 
via the long tail of the nuclear potential, the Slater determinant of the 
A particles by generating excited Slater's determinants. 
In this work the influence of the correlations on the level structure
and ground state distributions 
of even open shell nuclei is analyzed via
the boson dynamic correlation model BDCM. The model is based on
the unitary operator $e^S$ ({\it S} is the correlation operator) formalism which 
in this paper is presented within a non perturbative approximation. 
The low lying spectrum calculated for $^6$Li reproduce very well 
the experimental
spectrum while for $^6$He a charge radius slightly larger than that 
obtained within the isotopic-shift (IS) theory has been calculated.
Good agreement between theoretical and experimental results has 
been obtained without the introduction of a genuine three body force. 
}

\begin{document}

\maketitle

	     \section{Introduction}
Correlation effects in nuclei have been first introduced in nuclei by Villars~\cite{vil63},
who proposed the unitary-model operator (UMO) to construct effective operators.
The method was implemented by Shakin~\cite{sha66} for the calculation
of the G-matrix from hard-core interactions.
Non perturbative approximations of the UMO have been recently applied to even nuclei 
in Ref.~3) which here is treated in more detail. 
The basics formulas of the Boson Dynamic Correlation Model (BDCM) presented in the above
quoted paper have been obtained by solving the n-body problem in the following 
approximations:\\ 
a) The n-body correlation operator is separated in short- and long-range components.
 The short-range component is considered up to the two body correlation
while for the long range component 
the three and four body correlation operators have been studied.
The extension of the correlation operator to high order diagrams is especially important
in the description of exotic nuclei (open shell).  
In the short range approximation 
the model space of two interacting particles is separated in two subspaces:
one which includes the shell model states and the other (high momentum) which
 is used to compute the G-matrix of the model.
 The long range component of the correlation operator 
 has the effect of generating a new correlated model space (effective space)
which departs from the originally adopted one (shell model).
The amplitudes of the model wave functions are calculated in terms of
non linear equation of motions (EoM),\\
b) the n-body matrix elements are calculated exactly via the Cluster Factorization Theory 
(CFT),\\
c) by linearizing the systems of commutator equations,
which characterize the EoM. The generalized linearization approximations (GLA)
includes in the calculation presented in the paper up to the (3p1h) effective diagrams.
The linearized terms are not discarded but provide, as explained later in the text, the 
additional matrix elements that convert the perturbative UMO expansion 
in an eigenvalue equation. \\ 

Within the present treatment of the correlation operator one generates 
in the n-body theory not only the
ladder diagrams of Ref.~4) but also the folded diagrams of Kuo
~\cite{kuo01}.

In this paper the BDCM model is applied to calculate the influence of the 
correlations on the energy spectrum of $^6$Li and on the charge distributions of $^6$He
and $^6$Li.
The motivation of these calculations relays from one side
in the study the effect of the correlation operator on the theoretical charge radius
of $^6$He and from the other side in investigating the variation of
 the charge distribution and magnetic moment of the ground state of $^6$Li
under the variation of the adopted model space. 

 The value obtained for the charge radius of the correlated $^6$He is slightly
bigger than the radius calculated in other theories~\cite{nav98,pip01}
 and that derived within
the isotopic-shift IS theory~\cite{wan04}. A charge radius which agrees
 with the radii calculated in the Refs.~6,~9) and
those calculated in the cluster models of Refs.~11-13)
 is on the other hand obtained by considering
only two protons in the $1s_{\frac{1}{2}}$.
This non correlated radius agrees also with the radius derived at Argonne
within the IS theory~\cite{wan04}. 
Correlations have therefore the property to increase the charge radius
of $^6$He as observed
for the isotopes of Lithium.

 The calculations performed in Ref.~14) 
for the charge radii of the lithium isotopes, although in good agreement with those
 measured at GSI-TRIUMF~\cite{ewa05} and analyzed with the help of Ref.~16),
are always slightly larger than those measured.
For the stable isotope $^6$Li the calculated radius agrees with the value
obtained with the electron scattering experiments of  Ref.~17).
However, the charge radii calculated in the IS theory
could also depend on the nuclear correlations. 
The consideration of the microscopic correlations as presented in this paper 
will generate a new evaluation procedure for the Mass Shift (MS) and the Field Shift (FS). 
As result, both quantities could be evaluated within non-perturbative methods
which include the nuclear effects.   
 
The importance of the correlations in the evaluation of the FS has been
already pointed out in Ref.~19) where we calculate the FS of
$^7$Li and show that the departure from a point 
nuclear approximation is rather a big effect.   
Additionally the higher order cross term contributions of Ref.~18) 
need to be considered.  
A direct comparison between the calculated and the measured charge radii 
should be therefore performed after an accurate analysis of these two correcting factors.  

Theoretically the effect of the correlation on the distributions of medium-heavy nuclei
has been already performed in Refs~20-22) within a phenomenological correlation model.
In light nuclei, however the calculation of distributions performed 
by the theoretical models of Refs.~6,~7,~8,~9)
has been done in terms of non correlated particles.
In~Refs.~6,~7) and references therein quoted
large-basis no-core shell model have been used. In this model the $e^S$
method is considered up to the ``two body cluster approximation''. 
Within this approximation the effective interaction obtained 
contains no hole state. This will be however not the case by expanding
the $e^S$ method to higher cluster approximations as done in the present paper.     
 Unstable nuclei have been described by the antisymmetrized 
molecular dynamics (AMD) of Ref.~8) neglecting correlations.
Quantum Monte Carlo calculations have been performed by using realistic 
nuclear Hamiltonians that fit nucleon-nucleon scattering data
in light nuclei~\cite{pip01}. Good results have been obtained for structure calculations,
but the model can not introduce correlated wave functions in the calculation 
of the distributions.

Another important argument for the consideration of correlation effects
comes from the analysis of the magnetic moments.
The magnetic moment of the ground state of $^6$Li calculated in a model base
in which the hole is confined to the $1s_{\frac{1}{2}}$ is smaller than the experimental value.
Only with the use of a large configuration base, which includes the spin-flip
component $1p^{-1}_{\frac{3}{2}}1p_{\frac{1}{2}}$ proposed by Arima~\cite{ari02}
we obtain a very good value for the magnetic moment.
The large configuration base has also the effect of decreasing the energy of 
the second $1^-$ level.
\section{Theory of Correlated Two Particle Systems}
The effect of the correlations between nucleons in open shell nuclei is investigated within 
a system of coupled commutator equations, which via    
 the GLA Ansatz, is converted in an eigenvalue equation.
These describe a situation in which valence particles 
and core-excited states are coexisting.
 The advantage of this model is to 
provide larger effective configuration spaces (see Appendix A) than that used in 
the shell model calculation, 
to include exactly the Pauli principle in the coupled spaces, 
and to generate correlated solutions for the $n~\equiv$ paired systems.

We start by computing the commutator of the valence particles 
with the nuclear Hamilton's operator. In this calculation we retain the
linear (shell model) terms with dimension {\it n} and
non-linear (valence pairs coupled to one particle-hole pair) excitations with dimension $n+1'$.
The $1'$ denotes one particle-hole pair.
In the next equation we have then to take care of the 
non linear terms derived in the first step. 
The commutator of the Hamilton's operator with the $n+1'$ excited states 
generates the coupling of the
n-valence pairs to two particles-holes (2p-2h) excitations with dimension $n+2'$.
The successive equations are then characterizing the commutators which
involve valence particles coupled to an increasing number of particle-hole pairs $n+n'$.  
It is worthwhile to remark that the obtained system of commutator equations is similar to
the chain of equations one derives within the Green's function dynamics of Ref.~9).
The introduced computational steps describe the mixing of the shell
model states to core excitations with an increasing degree of complexity
 which will find applications in the calculation of the structures of exotic nuclei.
This commutator chain is suitable to be solved perturbatively by inserting 
the $n$-th commutator in the $(n-1)$-th commutator, $(n-1)$-th commutator 
in the $(n-2)$-th commutator 
, \ldots the second commutator in the first.
Within this perturbative approach one defines effective Hamiltonians 
of the model which, due to the increasing degree of complexity, are not easily solvable.
Much simpler solutions to the commutator equations may, however, be obtained in
the BDCM model.
We start by remarking that in the study of low lying excitations of the
n-body systems the higher order components of the wave functions, which involve 
 n valence - and (2p-2h) core-excitations are poorly admixed in the
model space and can be linearized. Within this approximation, the model
commutator equations are suitable to be restricted to a finite space. 
The linearized system of the commutator equations is then
solved exactly in terms of the CFT which calculates the n-body
matrix elements in an expedite and exact way.

In the following we illustrate the method by considering two valence particles.  
Following Ref.~2) we calculate the effective Hamiltonian by using only the $S_2$ 
correlation operator obtaining:
\begin{equation}\label{eq.1}
\begin{array}{l}
H_{eff}=e^{-iS_2}He^{iS_2}=\sum_{\alpha\beta}\langle\alpha|t|\beta\rangle 
a^{\dagger}_{\alpha}a_{\beta}+
\sum_{\alpha\beta\gamma\delta}\langle\Psi_{\alpha\beta}|v^l_{12}|\Psi_{\gamma\delta}\rangle a^{\dagger}_{\alpha}a^{\dagger}_{\beta} a_{\delta}a_{\gamma}\\
=\sum_{\alpha\beta}\langle\alpha|t|\beta\rangle a^{\dagger}_{\alpha}a_{\beta}+
\sum_{\alpha\beta\gamma\delta}\langle\Psi_{\alpha\beta}|v|\Psi_{\gamma\delta}\rangle a^{\dagger}_{\alpha}a^{\dagger}_{\beta} a_{\delta}a_{\gamma}
\end{array}
\end{equation}
where $v_{12}^l$ is the long component of the two body interaction (note that the
$v^l_{12}$ is in the following equations simply denoted as v).
 The $\Psi_{\alpha\beta}$ is the two particle correlated wave function:
\begin{equation}\label{eq.1a}
\Psi_{\alpha\beta}=e^{iS_2}\Phi_{\alpha\beta}
\end{equation}

In dealing with complex nuclei however the ($S_i,~i=3\cdots n$)
correlations should also be considered.

The evaluation of these diagrams is, due to the 
exponentially increasing number of terms, difficult in a perturbation theory.

We note however that one way to overcome this problem is to work with 
$e^{i(S_1+S_2+S_3+\cdots+S_i)}$ operator on the Slater's determinant
 by keeping the n-body Hamiltonian unvaried.

After having performed the diagonalization of the n-body Hamilton's operator we
can calculate the form of the effective Hamiltonian which, by now, includes 
correlation operators of complex order.

We write the two particle states in second quantization by discarding for
simplicity the isospin quantum numbers: 
\begin{equation}\label{e1}
\Phi_{2p}\longrightarrow A^{\dagger}_1(\alpha_1J)|0\rangle=[a^{\dagger}_{j_1}a^{\dagger}_{j_2}]^J_M|0\rangle,
\end{equation}
where the operators $a^{\dagger}_{j_1}a^{\dagger}_{j_2}$ create two coupled 
particles in the open shells and we analyze the structure of
the particle dynamics, generated by the correlation operator, via the following commutator:
\begin{equation}\label{e2}
\begin{array}{l}
\displaystyle{ 
[H,A^{\dagger}_1(\alpha_1J)]|0\rangle=
[(\sum_{\alpha}\epsilon_{\alpha}a^{\dagger}_{\alpha}a_{\alpha}+\frac{1}{2}
\sum_{\alpha \beta \gamma \delta} \langle\alpha\beta|v(r)|\gamma\delta\rangle a^{\dagger}_{\alpha} 
a^{\dagger}_{\beta}a_{\delta}a_{\gamma}),(a^{\dagger}_{j_1}a^{\dagger}_{j_2})^J]|0\rangle}.
\end{array}
\end{equation}
In order to have a compact index definition we have introduced:
\begin{equation}\label{e3}
\begin{array}{c}
\alpha_1\longrightarrow j_1j_2
\end{array}
\end{equation}
By using some operator's algebra and by including in the results
linear and nonlinear terms we calculate:
\begin{equation}\label{e4}
[H,A^{\dagger}_1(\alpha_1J)]|0\rangle=\sum_{\beta_{1}} \Omega(2p|2p') A^{\dagger}_1(\beta_1J)]|0\rangle+ 
\sum_{\beta_2J'_1J'_2} \Omega(2p|3p1h) A^{\dagger}_2(\beta_2J'_1J'_2J)]|0\rangle.
\end{equation}
In Eq.~(\ref{e4}) the $A^{\dagger}_1(\beta_1J)$ operators are those of Eq.~(\ref{e1}) and the 
 $A^{\dagger}_2(\beta_2J'_1J'_2J)$ are defined below:
\begin{equation}\label{e5}
\begin{array}{l}
\Phi_{3p1h}\longrightarrow A^{\dagger}_2(\beta_2J'_1J'_2J)|0\rangle
=((a_{j'_1}^{\dagger}a_{j'_2}^{\dagger})^{J'_1}(a_{j'_3}^{\dagger}a_{j'_4})^{J'_2})^J|0\rangle.
\end{array}
\end{equation}
In Eq.~(\ref{e5}) we have used the additional convention:
\begin{equation}\label{e6}
\begin{array}{c}
\beta_2\longrightarrow j'_1j'_2j'_3j'_4
\end{array}
\end{equation}
and we have associated:
 \begin{equation}\label{e7}
\begin{array}{cc}
J'_1~$to the coupling of$~j'_1j'_2\\
J'_2~$to the coupling of$~j'_3j'_4
\end{array}
\end{equation}
Having extended the commutator as in Eq.~(\ref{e4}), 
we have also to calculate the commutator equation
for the $A^{\dagger}_2(\alpha_2J_1J_2J)$ operators as given below: 
\begin{equation}\label{e8}
\begin{array}{l}
\displaystyle{
 [H,A^{\dagger}_2(\alpha_2J_1J_2J)]|0\rangle}  \\
\displaystyle{
=\sum_{\beta_{2}J'_1J'_2} \Omega(3p1h|3p'1h') A^{\dagger}_2(\beta_2J'_1J'_2J)|0\rangle+
\sum_{\beta_{3}J'_1J'_2J'_3} \Omega (3p1h|4p2h) A^{\dagger}_3(\beta_3J'_1J'_2J'_3J)|0\rangle},
\end{array}
\end{equation}
where we have introduced the (4p-2h) wave functions defined below:
\begin{equation}\label{e9}
\begin{array}{l}
\Phi_{4p2h}\longrightarrow A^{\dagger}_3(\beta_3J'_1J'_2J'_3J)|0\rangle=
(((a_{j'_1}^{\dagger}a_{j'_2}^{\dagger})^{J'_1}(a_{j'_3}^{\dagger}a_{j'_4})^{J'_2})^{J_{12}}(a_{j'_5}^{\dagger}
a_{j'_6})^{J'_3})^J|0\rangle,
\end{array}
\end{equation}
and where we have consistently extended the definition given in~(\ref{e3},\ref{e7}):
\begin{equation}\label{e10}
\begin{array}{c}
\beta_3\longrightarrow j'_1j'_2j'_3j'_4j'_5j'_6
\end{array}
\end{equation}
with:
 \begin{equation}\label{e11}
\begin{array}{cc}
J'_1~$associated to the coupling of$~j'_1j'_2\\
J'_2~$associated to the coupling of$~j'_3j'_4\\
J'_3~$associated to the coupling of$~j'_5j'_6
\end{array}
\end{equation}
In the definition of $A^{\dagger}_3(\beta_3J'_1J'_2J'_3J)$ the coupling of $J'_1$ to $J'_2$ to $J_{12}$
has been discarded from the notation.
In Eqs.~(\ref{e4},\ref{e8}) the $\Omega$'s are the matrix elements of the Hamilton's 
operator in the model wave functions.
The next step would be then the computation of the commutator
of the Hamiltonian with the $A^{\dagger}_3(\beta_3J'_1J'_2J'_3J)$ operators.
Here we linearize these contributions
by considering that in the study of the low energy spectrum and in the calculation of
 ground-state correlated distributions the $A^{\dagger}_3(\beta_3J'_1J'_2J'_3J)$ terms 
are poorly contributing.
The linearization is performed as in Ref.~25) by applying to the (4p2h) terms:
\begin{equation}\label{e12}
 \sum_{\alpha \beta \gamma \delta} \langle\alpha\beta|v(r)|\gamma\delta\rangle a^{\dagger}_{\alpha} 
a^{\dagger}_{\beta}a_{\delta}a_{\gamma}A^{\dagger}_3(\beta_3J'_1J'_2J'_3J)
\end{equation} 
the Wick's theorem and to discard the normal order terms.
Within this linearization approximation we generate from 
the commutator equations of Eq.~(\ref{e4},\ref{e8}) non perturbative solutions
of the EoM, i.e.: the eigenvalue equations for the mixed mode system:
\begin{equation}\label{e13}
\begin{array}{l}
\displaystyle{ 
[H,A^{\dagger}_1(\alpha_1J)]|0\rangle=
\sum_{\beta_{1}} \Omega(2p|2p') A^{\dagger}_1(\beta_1J)|0\rangle+ 
\sum_{\beta_2J'_1J'_2} \Omega(2p|3p'1h') A^{\dagger}_2(\beta_2J'_1J'_2J)|0\rangle},
\end{array}
\end{equation}
and
\begin{equation}\label{e13a}
\begin{array}{l}
\displaystyle{
 [H,A^{\dagger}_2(\alpha_2J_1J_2J)]|0\rangle}  
\displaystyle{
=\sum_{\beta_{1}} \Omega(3p1h|2p') A^{\dagger}_1(\beta_1J)|0\rangle+
\sum_{\beta_{2}J'_1J'_2} \Omega (3p1h|3p'1h') A^{\dagger}_2(\beta_2J'_1J'_2)|0\rangle}.
\end{array}
\end{equation}
Within the application of the GLA approximation we convert 
 Eqs.~(\ref{e4},\ref{e8}) in an eigenvalue equation for  
the configuration mixing wave functions (CMWFs) of the model.  
In fact, the linearization provides the additional matrix elements necessary 
to write the following identity:
\begin{equation}\label{e14}
\Omega(3p1h|3p'1h')= \langle j_1j_2j_3j_4|v(r)|j'_1j'_2j'_3j'_4\rangle,
\end{equation}
and to introduce the off-diagonal matrix elements 
which couple the (2p) to the (3p1h) subspaces.
Now, by writing
  Eqs.~(\ref{e13},\ref{e13a}) in the following matrix form:
\begin{equation}\label{e14a}
\begin{array}{l}
\left ( \begin{array}{c}
\displaystyle{
[H,A^{\dagger}_1(\alpha_1J)]|0\rangle} \\
\displaystyle{
[H,A^{\dagger}_2(\alpha_2J_1J_2J)]|0\rangle} \end{array} \right )
=\left ( \begin{array}{cc}
E_{2p}+\Omega(2p|2p')&\Omega(2p|3p'1h')\\
\Omega(3p1h|2p')&E_{3p1h}+\Omega(3p1h|3p'1h')\end{array} \right )
\left ( \begin{array}{c}
A^{\dagger}_1(\beta_1J)|0\rangle\\
A^{\dagger}_2(\beta_2J_1J_2J)|0\rangle \end{array} \right),
\end{array}
\end{equation}
and by multiplying to the left with:
\begin{equation}\label{e14b}
\begin{array}{l}
\left ( \begin{array}{c}
\langle0|A_1(\alpha_1J)\\
\langle0|A_2(\alpha_2J_1J_2J) \end{array} \right )
\end{array}
\end{equation}
we generate the eigenvalue equation for the dressed particles:
\begin{equation}\label{e15}
\begin{array}{l}
\displaystyle{\sum_{\beta_1\beta_2J'_1J'_2}
\left ( \begin{array}{cc}
E_{2p}+\langle A_1(\alpha_1J)|v(r)|A^{\dagger}_1(\beta_1J)\rangle & \langle A_1(\alpha_1J) |v(r)| A^{\dagger}_2(\beta_2J'_1J'_2J)\rangle\\
\langle A_2(\alpha_2J_1J_2J)|v(r)|A^{\dagger}_1(\beta_1J)\rangle 
& E_{3p1h}+ \langle A_2(\alpha_2J_1J_2J) |v(r)| A^{\dagger}_2(\beta_2J'_1J'_2J)\rangle \end{array} \right)}\\
 \cdot \left ( \begin{array}{c}  
 \chi_1(\beta_1J) \\
 \chi_2(\beta_2J'_1J'_2J)\end{array}\right)
=E \left (\begin{array}{c} 
 \chi_1(\alpha_1J)\\
 \chi_2(\alpha_2J_1J_2J)\end{array}\right )|0\rangle.
\end{array}
\end{equation}
In Eq.(\ref{e15}) 
$E_{2p}=\epsilon^{HF}_{j_1}+\epsilon^{HF}_{j_2}$ and
$E_{3p1h}=\epsilon^{HF}_{j_1}+\epsilon^{HF}_{j_2}+\epsilon^{HF}_{j_3}-\epsilon^{HF}_{j_4}$ 
are the Hartre-Fock energies (see Appendix C)
while the $\chi$'s are the projections of the model states:
\begin{equation}
\label{e15a}
|\Phi^J_{2p}\rangle=\chi_1(\alpha_1J)A_1(\alpha_1J)|0\rangle+\chi_2(\alpha_2J_1J_2J)A_2(\alpha_2J_1J_2J)|0\rangle
\end{equation}
to the basic vectors 2p,~3p1h.
To conclude, although the (4p-2h) CMWFs are not active part of  the 
model space, they are important for structure calculations.
On may therefore associate the GLA approximation to a parameter which
describes the degree of complexity of the model CMWFs
(the method used to define the CMWFs is given in the Appendices A, B, C).
Within the first order linearization we obtain the EoM for the shell model
while within the second and third order linearization approximations we derive
the EoM of valence particles coexisting with the complex particle-hole
structure of the excited states.    

In this paper we solve Eq.~(\ref{e15}) self-consistently (see Appendix D).
The solutions for the first iteration step are obtained by diagonalizing the 
eigenvalue equation~(\ref{e15}).
 The first step of the iterative method generates the dynamic amplitudes 
for the two dressed particles, i.e. two particles coexisting
with the 3p1h structures.  
With the calculated eigenvectors we recompute then the matrix elements 
$\langle j_1j_2|v(r)|j_1j_2j_3j_4\rangle$ and $\langle j_1j_2j_3j_4|v(r)|j'_1j'_2j'_3j'_4\rangle$
and we diagonalize again the eigenvalue equation.
The iterations are repeated until the stabilization of the energies has been
reached.

To be noted that since we are working in coupled particle-particle
and particle-hole bases we need both particle-particle and particle-hole matrix elements
(these terms where set to zero in the original work of Br{\"u}ckner~\cite{bru01})
in orders to diagonalize the eigenvalue equation.   
The calculation of these matrix elements is, however, complicated by the number of terms
which have to be evaluated in order to solve the introduced iterative equations.  
As in Ref.~3) one solves this problem by using the cluster factorization theory  
CFT which provides a quick and exact numerical method to perform the 
calculations of the matrix elements. 
The starting point of the CFT theory is to expand 
the CMWFs base introduced for (3p1h) states in terms of  
cluster factorization coefficients (CFC) denoted in the following U and V:  
\begin{equation}\label{e16}
\begin{array}{l}
\displaystyle{
|3p1h\rangle^J=A^{\dagger}_2(\alpha_2J_1J_2J)|0\rangle}\\
\displaystyle{
=\sum_{J_rJ_s\epsilon_i}{^5}V^{(3,1)}_J(\alpha_2J_1J_2|\}\epsilon_iJ_r\bar{\epsilon}_iJ_s)
|[A^{\dagger}_1(\epsilon_iJ_r)B^{\dagger}_1(\bar{\epsilon}_iJ_s)]^J}|0\rangle\\
\displaystyle{
+\sum_{J_rJ_s\alpha_j}
{^5}U^{(3,1)}_J(\alpha_2J_1J_2|\}\alpha_jJ_r\bar{\alpha}_jJ_s)
|[B^{\dagger}(\alpha_jJ_r)A^{\dagger}_1(\bar{\alpha}_jJ_s)]^J|0\rangle}\\
\displaystyle{
=\sum_{J_rJ_s\epsilon_i}{^5}V^{(3,1)}_J(\alpha_2J_1J_2|\}\epsilon_iJ_r\bar{\epsilon}_iJ_s)|[\epsilon_iJ_r\bar{\epsilon}_iJ_s]^J}
|0\rangle\\
\displaystyle{
+\sum_{J_rJ_s\alpha_j}
{^5}U^{(3,1)}_J(\alpha_2J_1J_2|\}\alpha_jJ_r\bar{\alpha}_jJ_s)|[\alpha_jJ_r\bar{\alpha}_jJ_s]^J|0\rangle},
\end{array}
\end{equation}
where the $(\alpha_2)$ coordinates of the (3p1h) model states have been expanded
in terms of active (passive) particle-particle and passive (active) 
particle-hole coordinates as given below:
\begin{equation}\label{e17}
\begin{array}{l}
\begin{array}{c}
\alpha_2\to\epsilon_i(J_r)$(active~particle-particle)$\bar{\epsilon}_i(J_s)
$(passive~particle-hole)$\\
+\alpha_j(J_r)$(active~particle-hole)$\bar{\alpha}_j(J_s)
$(passive~particle-particle)$
\end{array}
\end{array}
\end{equation}
In Eq.~(\ref{e16}) the over-script on the left of the V and U 
indicates the total number of pairs
(2(two pairs)+1=5) while the over-scripts on the right the number of particles and holes.
 There we have also introduced the creation operators of a particle-hole pair:
\begin{equation}\label{e18}
(ph)\longrightarrow B^{\dagger}_1(\alpha_1J)|0\rangle=[a^{\dagger}_{j_1}a_{j_2}]^J_M|0\rangle 
\end{equation}
 and postulated that the sum over (i) is to be extended over the six 
combinations (partitions)
of particle-particle pairs formed by the (3p1h) model space and
(j) over the three combinations of particle-hole pairs  
\begin{equation}\label{e19}
\begin{array}{ccc}
\alpha_2\to& \epsilon_i\bar{\epsilon}_i & \alpha_j\bar{\alpha}_j\\ \hline
j_1j_2j_3j^{-1}_4\to& (j_ij_j)(j_kj^{-1}_4) &(j_kj^{-1}_4)(j_ij_j)
\end{array}
\end{equation}
Within this convention we reproduce exactly the 
the (3p1h) CMWFs. 
In Eq.~(\ref{e16}) the sum over $(\bar{\alpha}_i)$ and $(\bar{\epsilon}_i)$ 
has not been given explicitly because these indices are complementary to the 
$(\alpha_i)$ and $(\epsilon_i)$ in the sense of Eq.~(\ref{e19}).
The sum over $J_r,J_s$ take care of the fact
that by calculating the (3p1h) matrix elements in the traditional way,
 the coupling of the j's of the (2p) is not the same as in the
 coupling postulated for (3p1h) wave functions. 
The definition of active and passive components
separate the matrix elements of the interaction 
in particle-particle and particle-hole matrix elements.

It has to be remarked that, due to the used iterative method, 
 the final results are almost independent from the 
  initial choice of the interaction because after the first
iteration the two body potential is modified by the contributions 
 of both types of matrix elements. 

Matrix elements calculated by using the first term of Eq.~(\ref{e16}) are
 related to the Br{\"u}ckner theory~\cite{bru01},
 while those evaluated with the second term have been considered 
in the folded diagram theory of Kuo~\cite{kuo01}.

In order to calculate the CFC-${^5}V^{(3,1)}_J(\alpha_2J_1J_2|\}\epsilon_iJ_r\bar{\epsilon}_iJ_s)$ 
we introduce~\cite{tom01} the following operator:
\begin{equation}\label{e20}
\pi^k_{m_k}(2)=[(a^{\dagger}_{i'}a^{\dagger}_{i})^{J_i}(a_{j} a_{j'})^{J'_i}]^k_{m_k}
\end{equation}
 which destroys and creates a particle pair 
and we evaluate his effect on the $3p1h$ wave-function. 
In Eq.~(\ref{e16}) the two in the parenthesis indicates that the operators
 are working on the space spanned by two boson (two pairs). We calculate:
\begin{equation}\label{e21}
\begin{array}{l}
\displaystyle{
\pi^k_{m_k}(2)A^{\dagger}_2(\alpha_2J_1J_2JM)|0\rangle= C^{JkJ}_{Mm_kM'} }
($Coef$_1\cdot([a^{\dagger}_{i'}a^{\dagger}_i]^{J_r}[a^{\dagger}_{j_3}a_{j_4}]^{J_s})^J_{M'}\\ +
$Coef$_2\cdot([a^{\dagger}_ia^{\dagger}_{j_2}]^{J_r}[a^{\dagger}_{i'}a_{j_4}]^{J_s})^J_{M'}   +
$Coef$_3\cdot([a^{\dagger}_{j_1}a^{\dagger}_i]^{J_r}[a^{\dagger}_{i'}a_{j_4}]^{J_s})^J_{M'})|0\rangle.
\end{array}
\end{equation}
The Coef$_{\lambda_i},\lambda_i=1,2,3$ are given in the appendix E.
By multiplying Eq.~(\ref{e21}) to the left with $\langle0|A_2(\alpha_2J_1J_2)$ we obtain 
the following equation:
\begin{equation}\label{e22}
\begin{array}{l}
\frac{\langle0|A_2(\alpha_2J_1J_2JM)\pi^k_{m_k}(2)A^{\dagger}_2(\alpha_2J_1J_2JM)|0\rangle}
{(\mathrm{Coef}_1+\mathrm{Coef}_2+\mathrm{Coef}_3)}
=C^{JkJ}_{Mm_kM'}
\end{array}
\end{equation}
where we have introduced the Clebsch-Gordan coefficients $C^{JkJ}_{Mm_kM'}$.
Eq.~(\ref{e22}) defines the matrix elements of a unitary operator, i.e.: by
 applying the unitary operator 
$\frac{{\pi}^k_{m_k}(2)}{\mathrm{Coef}_1+\mathrm{Coef}_2+\mathrm{Coef}_3}$
on the left of a wave function of the type $|3p1h\rangle^J_M$ we reproduce  
the same wave function in a rotated frame.
It follows that the structure defined by Eq.~(\ref{e22}) is that of a full linear group in
(2J+1) dimension and of its unitary subgroup $U_{2J+1}$.
The calculation of the transformation coefficients is now reduced to 
the construction of the Coef$_{\lambda_i}$ coefficients.
In order to demonstrate this,
let us now operate with the Casimir's operator of the group 
$(\pi^k_{m_k}(2)\otimes (\pi^k_{m_k}(2))^{\dagger})^0_0$ on the
left side of Eq.~(\ref{e16}). We obtain:
\begin{equation}\label{e23}
\langle A_2(\alpha_2J_1J_2J)\| (\pi^k_{m_k}(2) \otimes (\pi^k_{m_k}(2))^{\dagger})^0_0\|A^{\dagger}_2(\alpha_2J_1J_2J)\rangle
=\sum_{\lambda_i\lambda_j}\mathrm{Coef}_{\lambda_i}\mathrm{Coef}_{\lambda_j}
\end{equation}
where the double bar matrix elements have been introduced because 
the CFC are independent from the m's quantum numbers and where the subindex $\lambda_i$
classifies the three different partitions spanned by the particles in the
(3p1h) wave functions.
On the other hand considering that: 
\begin{equation}\label{e24}
(\pi^k_{m_k}(2)\otimes (\pi^k_{m_k}(2))^{\dagger})^0_0=
\sum_{i=1}^{\lambda_i}(\pi^k_{m_k}(1,i)\pi^0_0(\bar1,i)\otimes((\pi^0_0(\bar1,i))^{\dagger}(\pi^k_{m_k}(1,i))^{\dagger})^0_0
\end {equation}
where the sum is running over all the possible partitions, we calculate
by using Eq.~(\ref{e16}):
\begin{equation}\label{e25}
\begin{array}{l}
\langle A_2(\alpha_2J_1J_2J)||(\pi^k_{m_k}(1,i)\pi^0_0(\bar1,i)\otimes((\pi^0_0(\bar1,i))^{\dagger}(\pi^k_{m_k}(1,i))^{\dagger})^0_0
||A^{\dagger}_2(\alpha_2J_1J_2J)\rangle\\
={^5}V^{\dagger(3,1)}_J(\alpha_2J_1J_2|\}\lambda_i\epsilon_iJ_r\bar{\epsilon_i}J_s)
{^5}V^{(3,1)}_J(\alpha_2J_1J_2|\}\lambda_i\epsilon_iJ_r\bar{\epsilon_i}J_s).
\end{array}
\end{equation}
In Eq.~(\ref{e25}) we have introduced the unit operators $\pi^0_0(\bar1,i)$ defined by:
\begin{equation}\label{e26}
\langle B_1(\bar{\epsilon}_jJ'_s)|(\pi^0_0(\bar1,i))^{\dagger}\pi^0_0(\bar1,i)|B^{\dagger}_1(\bar{\epsilon}_iJ_s)\rangle=\delta_{\bar{\epsilon}_i\bar{\epsilon}_j}\delta_{J_sJ'_s}.
\end{equation} 
By equating Eq.~(\ref{e23}) and Eq.~(\ref{e25}) we obtain:
\begin{equation}\label{e27}
{^5}V^{\dagger(3,1)}_J(\alpha_2J_1J_2|\}\lambda_i\epsilon_iJ_r\bar{\epsilon_i}J_s)
{^5}V^{(3,1)}_J(\alpha_2J_1J_2|\}\lambda_i\epsilon_iJ_r\bar{\epsilon_i}J_s)= 
\sum_{\lambda_i\lambda_j}\mathrm{Coef}_{\lambda_i}\mathrm{Coef}_{\lambda_j}.
\end{equation}
Eq.~(\ref{e27}) is given in matrix form in the Appendix E.
Note that from the diagonalization of the Eq.~(\ref{e35}) we derive the 
square of the CFC. In order to calculate the CFC we introduce the same phase
convention used to define the Clebsch-Gordan coefficients. 
Analogously also for the operator  
\begin{equation}\label{e28}
u^k_{m_k}(2)=
[(a^{\dagger}_{i}a_{j})^{J_i}(a^{\dagger}_{j'} a_{i'})^{J'_i}]^k_{m_k}
\end{equation}
which destroys and creates a particle-hole pair.
 The normalization factors for these operators $(c_{\mu_i};i=4,5,6)$ are given 
in the appendix D.
A numerical example for the calculation of these coefficients is given in Appendix F.
By using Eq.~(\ref{e35}) and the evaluated CFC, we can now calculate the matrix elements of the
nuclear interaction in the (3p1h) CMWFs. 
We derive:
\begin{equation}\label{e29}
\begin{array}{l}
\displaystyle{
\langle3p1h|v(r)|3p'1h'\rangle= \langle A_2(\alpha_2J_1J_2J)|v(r)|A^{\dagger}(\beta_2J'_1J'_2J)\rangle}\\
\displaystyle{
=\sum_{\lambda_i\lambda_{j}\epsilon_i\epsilon_jJ_rJ_sJ'_rJ'_s}{^5}V^{\dagger}_J(\alpha_2J_1J_2|\}\lambda_i\epsilon_iJ_r\bar{\epsilon}_iJ_s)
{^5}V_J(\beta_2J'_1J'_2|\}\lambda_j\epsilon_jJ'_r\bar{\epsilon_j}J'_s)}\\
\displaystyle{
\langle[\lambda_i\epsilon_iJ_r\bar{\epsilon}_iJ_s]^J|v(r)|[\lambda_j\epsilon_jJ'_r\bar{\epsilon}_jJ'_s]^J\rangle}\\
\displaystyle{
+ \sum_{\mu_i\mu_j\alpha_i\alpha_jJ_rJ_sJ'_rJ'_s}{^5}U^{\dagger}_J(\alpha_2J_1J_2|\}\mu_i\alpha_iJ_r\bar{\alpha}_iJ_s)
{^5}U_J(\beta_2J'_1J'_2|\}\mu_j\alpha_jJ'_r\bar{\alpha}_jJ'_s)}\\
\displaystyle{
\langle[\mu_i\alpha_iJ_r\bar{\alpha}_iJ_s]^J|v(r)|[\mu_j\alpha_jJ'_r\bar{\alpha}_jJ'_s]^J\rangle}\\
\displaystyle{
=\sum_{\lambda_i\lambda_j\epsilon_i\epsilon_jJ_rJ'_rJ_s}{^5}V^{\dagger}_J(\alpha_2J_1J_2|\}\lambda_i\epsilon_iJ_r\bar{\epsilon}_iJ_s)
{^5}V_J(\beta_2J'_1J'_2|\}\lambda_j\epsilon_jJ'_r\bar{\epsilon}_jJ_s)}\\
\displaystyle{
\langle\lambda_i\epsilon_i|v(r)|\lambda_j\epsilon_j\rangle^{J_r}_a}\\
\displaystyle{
+ \sum_{\mu_i\mu_j\alpha_i\alpha_jJ_rJ'_rJ_s}{^5}U^{\dagger}_J(\alpha_2J_1J_2|\}\mu_i\alpha_iJ_r\bar{\alpha}_iJ_s)
{^5}U_J(\beta_2J'_1J'_2|\}\mu_j\alpha_jJ'_r\bar{\alpha}_jJ_s)}\\
\displaystyle{
\langle\mu_i\alpha_i|v(r)|\mu_j\alpha_j\rangle^{J_r}_a}.
\end{array}
\end{equation}
In Eq.~(\ref{e29}) we have two types of matrix elements, the 
particle-particle matrix elements which, by choosing one $\lambda_i$ and
$\mu_i$ partitions, can be written as following:
\begin{equation}\label{e30}
\begin{array}{l}
\langle\lambda_1\epsilon_1|v(r)|\lambda'_1\epsilon'_1\rangle^{J_r}_a
=\langle j_1j_2|v(r)|j'_1j'_2\rangle^{J_r}_a,
\end{array}
\end{equation}
and the particle-hole matrix elements:
\begin{equation}\label{e31}
\begin{array}{l}
\langle\mu_1\alpha_1|v(r)|\mu'_1\alpha'_1\rangle^{J_r}_a
=\langle j_3j_4|v(r)|j_3j_4\rangle^{J_r}_a.
\end{array}
\end{equation}
The matrix elements of Eq.~(\ref{e31}) which generate the interaction with the core clusters
 are not considered by the microscopic
calculation of Refs.~6,~9) where the effective Hamiltonian is
obtained by summing only over ladder diagrams \cite{bru01}.
By using the diagonal matrix elements of Eq.~(\ref{e29}) 
and the off-diagonal matrix elements given in Ref.~3) we diagonalize Eq.~(\ref{e15})
and obtain the amplitudes $\chi(\alpha_1)$ and $\chi(\alpha_2)$ of the mixed modes
(2p) and (3p1h).
The calculation of the ground and excited state distributions and 
of the magnetic moments is then performed in terms of these amplitudes.
The formula of the correlated distribution for the ground state of
even nuclei is given in Appendix H.
\section{Results}
In order to perform structure calculations, we have to define a single-particle
base with the relative ``single-particle energies'' and to choose the
 nuclear two-body interactions.
The single-particle energies of these levels are taken from the known experimental
level spectra of the neighboring nuclei~\cite{til01} (see Appendix C) and given in Table~I. 
For the experimentally unknown single particle energies of the fp shells we use the
corresponding energies for the mass A-9 nuclei scaled accordingly the
different binding energies.
Our single particle energies agree reasonably well with those calculated
in Ref.~28).
Some of these levels are not bound. 
In this paper we perform as in Ref.~9) calculations by assuming all levels as bound.
It has however to be remember that the energies of Table~I are used only
in the first stage of the iteration procedure explained in Appendix F). 

For the particle-particle interaction, we use the G-matrix obtained from
 Yale potential~\cite{sha67}.
These matrix elements are evaluated by applying the
$e^S$ correlation operator, truncated at the second order
term of the expansion, to the harmonic oscillator
base with size parameter b=1.76 fm.   
This value is consistent with the value used by Kuo~\cite{kuo02}.
As elucidate in Ref.~3) the potential used by the BDCM 
is separated in low and high momentum components.
 Therefore, the effective model matrix elements calculated within 
 the present separation method
and those calculated by Kuo~\cite{jia01} are pretty similar.
The separation method generates matrix elements which are almost
independent from the radial shape of the different potentials generally
used in structure calculations. \\
The particle-hole matrix elements could be calculated from the particle-particle 
matrix elements via a re-coupling transformation. 
We prefer to use the phenomenological potential of Ref.~24). 
The same size parameter as for the particle-particle
 matrix elements has been used.

One can generate the center-of-mass (CM) spurious states according to Refs.~32,~33)
and evaluate the overlap between these states and the nuclear
eigenstates of the model (see Appendix B).
Model components having with the corresponding CM components an overlap greater than 10\% 
were treated as spurious states and discarded.
This is a convenient approximation considering that in our model space the energy 
of the CM is varying between 18 and 20 MeV.

By using a base formed by 9 (2p) and 112 3p1h-states we calculate
the distribution of the ground state of $^6$He given in Fig.~\ref{fig.1}.
The spectroscopic factor for the ground state wave function, defined in Appendix G,
and the most significant components of the ground state wave function are given
 in Table IX. 

In Fig.~\ref{fig.1} we plot three distributions: 
1) the correlated charge distribution of $^6$He 
calculated with Eq.~(\ref{e46}), 
2) the correlated charge distribution of $^6$He 
calculated with Eq.~(\ref{e46}) but neglecting the folded diagrams, 
3) the charge distribution calculated 
for two correlated protons in the $1s_{\frac{1}{2}}$ shell. 
A charge radius of 2.25 fm has been obtained for the distribution a), a radius of 2.39 fm
for the distribution b), and a radius of 2.09 fm for the distribution c).

In Table~(II) the calculated charge radii are compared with the radii calculated by 
the other theoretical models and with the radius obtained by the IS theory.
The radius calculated with the distribution c) (single particle) is in agreement with
the radius obtained by the IS theory and with those of the other theoretical models. 
The BDCM increases however the charge radius of $^6$He.
One reason for this slightly disagreement can relay on the
 effect of the correlation operator which is neglected either in the 
 evaluation of the charge radii via the Is theory or in the other theoretical model 
calculations.
As shown in Ref.~3) the correlations are also important
in the analysis of the matter radius of $^6$He.
Here the larger matter radius reproduces reasonable well the
proton elastic scattering cross section measured at GSI at 717~MeV/u calculated by
using the Glauber method in the whole range of data~\cite{aks02}.
  
The energy of the low lying spectrum of $^6$Li is calculated by defining
a coupled base in which a proton-neutron pair in the (psd) single particle scheme of Table~(I)
are interacting with the (3p1h) states generated by exciting the hole from the lowest 
(1s) shell. Within the model dimension of 13 (2p) and  196 (3p1h) components
we define the ground state wave functions. The spectroscopic factor for the ground state 
of $^6$Li and the most significant components of the
ground state wave function are given in Table X.  
By using this model space we obtain the spectrum shown in Fig.~\ref{fig.2} right. 
For the ground state this model space is 
however not large enough to reproduce the 
second $1^+$ level and the magnetic moment of the ground state~\cite{ajz84}  .
The magnetic moment of the ground state calculated in this model space 
is 0.875 nm therefore larger then the experimental value (0.822 nm).
The second spectrum is calculated
introducing for the ground state a much larger space with dimension of 525 components 
 (13 two particle shell model states and 512 (3p1h) states) 
which includes also the excitation of 
the particles from the $1p_\frac{3}{2}$ state. 
Within this dimensional space the spectroscopic factor for the ground state 
of $^6$Li, defined in appendix G, and the most significant components are
given in Table XI.  
Within this space we reproduce also the energy of 
the second $1^+$ level given in Fig.~\ref{fig.2} left.
To be considered that in other theoretical models the $1^+$ is always lying
at too high energy.
Since we plot the spectrum relative to the ground state energy,
the effect of enlarging the base also for the other states
is negligible at least for the first plotted $2^+$ and $3^+$ levels. 
The magnetic moment of the ground state calculated for the large model space 
is 0.821 nm, value that is in good agreement with the experimental value.
The charge distributions calculated within these two model spaces is given
in Fig.~\ref{fig.3}. The two distribution are labeled: 1) charge distribution calculated for
the small configuration space and 2) charge distribution calculated for the large space. 
The calculated charge radii are respectively $\langle r^2_{ch}\rangle^{\frac{1}{2}}$
equal to 2.627 fm for the small model space and to 2.55 fm  
 for the large space.
Both values reproduce well the charge radius of 2.55 fm obtained 
in Ref.~17) from the electron scattering experiments.
Calculation of the electric scattering form factor measured in Ref.~17) 
for both charge distributions is presently under consideration.
\section{Conclusions}
In this paper we have investigated the effect of the microscopic correlation operators 
on the spectra and the charge distributions of $^6$He and $^6$Li.
The microscopic correlation has been separated in short- and long-range
correlations according the definition of Shakin~\cite{sha66}.
The short-range correlation has been used to define the effective Hamiltonian
of the model while the long-range is used to calculate the structures
and the distributions of exotic nuclei.
As given in the work of Shakin, only the two body short range correlation 
need to be considered in order to derive the effective Hamiltonian
especially if the correlation is of very short range.
For the long range correlation operator however the ``three body component'' of
the correlation operator is important and should not be neglected.
The effect of the ``three body correlation operator'' is to introduce in the theory a three body
interaction. Therefore the use of the genuine three body interaction
of the other theoretical model could, in the present theory,
generate double counting of diagrams. \\
By using generalized linearization approximations and cluster factorization
 coefficients we can perform expedite and exact calculations for the structure
of $^6$He and $^6$Li.  
Very good results for the spectrum of $^6$Li have been obtained by considering 
large configuration spaces. 
Within correlated distributions we obtain charge radii slightly larger than those
calculated within either non correlated distributions or IS experiments. \\   
This result should serve as motivation to a reevaluation of the different
terms of the IS theory: i.e. the MS, which is mainly calculated in a perturbative
approximation, and the FS, which is generally calculated in a point nucleus approximation,
should be reevaluated within the non perturbative BDCM. 
\appendix
\section{Definition of the model CMWFs}
In the BDCM the degree of linearization applied to the commutator equations
defines the CMWFs of the model.
For A=6 the model space is formed by two valence particle states 
and by the full set of the (3p1h) CMWFs. 
These different components are associated to the following linearization 
mechanism:
a) In the zero order linearization approximation we retain
only two particle states:
\begin{equation}
\Psi^{2p}(j_1j_2J)=[a^{\dagger}_{j_1}a^{\dagger}_{j_2}]^{JM}|0\rangle
\end{equation}
For the two particles we distinguish between :\\
1) effective valence space which is used to diagonalize the EoM,\\
2) complementary high excited single particle states which are used to compute
the G matrix.\\
b) In the first order linearization approximation we include
in the dynamic theory also the (3p1h) terms.
 These are  generated by the
application of the correlation operator of the third order to the particles
in the open shell states.
Within this linearization approximation the CMWFs of the model are defined by:
\begin{equation}
\Psi^{dressed}(j_1j_2J)=([a^{\dagger}_{j_1}a^{\dagger}_{j_2}]+
 [a^{\dagger}_{j_1}a^{\dagger}_{j_2}]^{J_{12}}[a^{\dagger}_{j_3}a_{j_4}]^{J_{34}})^{JM}|0\rangle.
\end{equation}
The (3p1h) CMWFs are then expanded according to Eq.~(\ref{e16}). This expansion
allows to orthogonalize the CMWFs in an easy way. 
c) The (4p2h) states which characterize the second order linearization step
are not included in the model space but, linearized, generate
the eigenvalue equation of the model (2p)+(3p1h) states. 
\section{Center of mass correction} 
Before performing the diagonalization of relative Hamilton's operator in the 
CMWFs defined in appendix A) we have to eliminate the spurious center of 
mass states.
We start to compute, following the calculations of Refs.~32,~33),
the percent weights of spurious states in the model wave functions.
These  can be obtained by calculating the energy of the center of mass
according to the following equation:
\begin{equation}\label{c1}
\begin{array}{l}
\displaystyle{
E_R= \int dR\Psi^{\dagger dressed}(j_ij_jJ)(R^2)\Psi^{dressed}(j_ij_jJ)}\\
\displaystyle{
+2\sum_{ij} \int d\vec{r_i}d\vec{r_j}\Psi^{\dagger dressed}(j_ij_jJ)(\vec{r_i}\cdot\vec{r_j})\Psi^{dressed}(j'_ij'_jJ)}.
\end{array}
\end{equation}
In Eq.~(\ref{c1}) 
the calculation of the integrals can be performed by using the expansion
for the (3p1h) states given in Eq.~(\ref{e16}) and by considering that 
for two particle states we have:
\begin{equation} 
\begin{array}{l}
\langle j_ij_jJ|(\vec{r_i}\cdot\vec{r_j})|j_ij_jJ\rangle\\
=\frac{4\pi}{3}[\hat{j_i} \hat{j_j}]
\left ( \begin{array} {ccc}
 j_i & 1 &j_j\\
-\frac{1}{2}& 0 &\frac{1}{2} \end{array} \right )^2
\left \{
 \begin{array} {ccc}
j_i & j_j &J\\
j_i & j_j &1 \end{array} \right \}
\langle l_i|r|l_j\rangle^2,
\end{array}
\end{equation}
where:
\begin{equation}
\hat{j}=(2j+1).
\end{equation}
By diagonalizing the above operator in the model space we obtain the energy 
of the center of mass.
The overlap with the model space give
 the degree of ``spuriosity'' of the different components.
\section{The single particle energies}
The model space which characterize the BDCM is formed by
adding even particle to a closed-shell nucleus.
The closed shell configuration can be described by a single Slater determinant and one
can use the Hartree-Fock's theory to obtain the binding energy and the single-particle 
energies. Alternatively one can remark that for a closed shell nucleus (Z,N)
 the single particle energies for the states above the Fermi surface are related to
the binding energies differences:
\begin{equation}
\epsilon^>_p=BE(Z,N)-BE^*(Z+1,N),
\end{equation}
and
\begin{equation}
\epsilon^>_n=BE(Z,N)-BE^*(Z,N+1).
\end{equation}
The single particle energies for the states below the Fermi surface are given by:
\begin{equation}
\epsilon^<_p=BE^*(Z-1,N)-BE(Z,N),
\end{equation}
and
\begin{equation}
\epsilon^<_n=BE^*(Z,N-1)-BE(Z,N).
\end{equation}
The BE are ground states binding energies which are taken as positive values,
 and e will be negative for bound states. $(BE^*=BE-E_x)$ is the ground state 
binding energy minus the excitation energy of the excited states associated 
with the single particle states.
Within this method, which recently has been reintroduced by B.A. Brown~\cite{bro01},
we derive the single particle energies from the 
known spectra of neighbor nuclei (see Table  I).
\section{Iteration procedure to calculate dressed eigenstates}
Let us suppose we have two particles in the sd shell model states interacting via the
(3p1h) CMWFs. The  configuration mixed wave functions are:
\begin{equation}
\begin{array}{l}
\Psi_{2p}^{\mathrm{dressed}}=[a^{\dagger}_{1d_{\frac{5}{2}}}a^{\dagger}_{2s_{\frac{1}{2}}}]^2
+[a^{\dagger}_{1d_{\frac{3}{2}}}a^{\dagger}_{2s_{\frac{1}{2}}}]^2\\
\displaystyle{
+\left [ [a^{\dagger}_{1d_{\frac{5}{2}}}a^{\dagger}_{2s_{\frac{1}{2}}}]^3
[a^{\dagger}_{2s_{\frac{1}{2}}}a_{1s_{\frac{1}{2}}}]^1 \right ]^2}
\displaystyle{
+\left [ [a^{\dagger}_{1d_{\frac{3}{2}}}a^{\dagger}_{2s_{\frac{1}{2}}}]^2
[a^{\dagger}_{1d_{\frac{5}{2}}}a_{1s_{\frac{1}{2}}}]^3 \right ]^2}\\
= \psi_1+\psi_2+ \psi_3+\psi_4
\end{array}
\end{equation}
solution of the eigenvalue matrix:
\begin{equation}\label{m.1}
\begin{array}{l}
\left ( \begin{array}{ccccc}
E_1+ \langle\psi_1|v(r)|\psi_1\rangle & \langle\psi_1|v(r)|\psi_2\rangle & \langle\psi_1|v(r)|\psi_3\rangle & \langle\psi_1|v(r)|\psi_4\rangle \\
 \langle\psi_2|v(r)|\psi_1\rangle & E_2+\langle\psi_2|v(r)|\psi_2\rangle & \langle\psi_2|v(r)|\psi_3\rangle & \langle\psi_2|v(r)|\psi_4\rangle \\
 \langle\psi_3|v(r)|\psi_1\rangle & \langle\psi_3|v(r)|\psi_2\rangle & E_3+\langle\psi_3|v(r)|\psi_3\rangle & \langle\psi_3|v(r)|\psi_4\rangle \\
 \langle\psi_4|v(r)|\psi_1\rangle & \langle\psi_4|v(r)|\psi_2\rangle & \langle\psi_4|v(r)|\psi_3\rangle & E_4+\langle\psi_4|v(r)|\psi_4\rangle 
\end{array} \right )\\
\cdot \left | \begin{array}{c}
 \psi_1|0\rangle \\ \psi_2|0\rangle \\ \psi_3 |0\rangle \\ \psi_4|0\rangle \end{array} \right )=0
\end{array}
\end{equation}
By diagonalizing the matrix of Eq.~(\ref{m.1})
we obtain four eigenvalues $\tilde{E_1},\tilde{E_2},\tilde{E_3},\tilde{E_4}$
and the four eigenvectors given below:
\begin{equation}\label{c.1}
\tilde{\Psi_1}=\chi^1_1|\psi_1\rangle+\chi^1_2|\psi_2\rangle+\chi^1_3|\psi_3\rangle+\chi^1_4|\psi_4\rangle
\end{equation}
\begin{equation}
\tilde{\Psi_2}=\chi^2_1|\psi_1\rangle+\chi^2_2|\psi_2\rangle+\chi^2_3|\psi_3\rangle+\chi^2_4|\psi_4\rangle
\end{equation}
\begin{equation}
\tilde{\Psi_3}=\chi_1^3|\psi_1\rangle+\chi^3_2|\psi_2\rangle+\chi^3_3|\psi_3\rangle+\chi^3_4|\psi_4\rangle
\end{equation}
\begin{equation}
\tilde{\Psi_4}=\chi^4_1|\psi_1\rangle+\chi^4_2|\psi_2\rangle+\chi^4_3|\psi_3\rangle+\chi^4_4|\psi_4\rangle
\end{equation}
This eigenvalues and eigenvectors are then used to diagonalize the eigenvalue matrix
in the second iteration step: 
\begin{equation}\label{m.2}
\begin{array}{l}
\left ( \begin{array}{ccccc}
 \tilde{E_1}+ \langle\tilde{\Psi_1}|v(r)|\tilde{\Psi_1}\rangle & \langle\tilde{\Psi_1}|v(r)|\tilde{\Psi_2}\rangle & 
 \langle\tilde{\Psi_1}|v(r)|\tilde{\Psi_3}\rangle & \langle\tilde{\Psi_1}|v(r)|\tilde{\Psi_4} \rangle \\
 \langle\tilde{\Psi_2}|v(r)|\tilde{\Psi_1}\rangle & \tilde{E_2}+\langle\tilde{\Psi_2}|v(r)|\tilde{\Psi_2}\rangle & 
 \langle\tilde{\Psi_3}|v(r)|\tilde{\Psi_3}\rangle & \langle\tilde{\Psi_3}|v(r)|\tilde{\Psi_4}\rangle \\
 \langle\tilde{\Psi_3}|v(r)|\tilde{\Psi_1}\rangle & \langle\tilde{\Psi_3}|v(r)|\tilde{\Psi_2}\rangle & 
 \tilde{E_3}+\langle\tilde{\Psi_3}|v(r)|\tilde{\Psi_3}\rangle & \langle\tilde{\Psi_3}|v(r)|\tilde{\Psi_4}\rangle \\
 \langle\tilde{\Psi_4}|v(r)|\tilde{\Psi_1}\rangle & \langle\tilde{\Psi_4}|v(r)|\tilde{\Psi_2}\rangle &
 \langle\tilde{\Psi_4}|v(r)|\tilde{\Psi_3}\rangle & \tilde{E_4}+\langle\tilde{\Psi_4}|v(r)|\tilde{\Psi_4}\rangle 
\end{array} \right )=0
\end{array}
\end{equation}
where:
\begin{equation}\label{m.3}
\begin{array}{l}
\langle\tilde{\Psi_1}|v(r)|\tilde{\Psi_1}\rangle\\
=(\chi^1_1)^2\langle\psi_1|v(r)|\psi_1\rangle+(\chi^1_2)^2\langle\psi_2|v(r)|\psi_2\rangle+(\chi^1_3)^2\langle\psi_3|v(r)|\psi_3\rangle
+(\chi^1_4)^2\langle\psi_4|v(r)|\psi_4\rangle       \\
+2 \chi^1_1\chi^1_2\langle\psi_1|v(r)|\psi_2\rangle+2 \chi^1_1\chi^1_3\langle\psi_1|v(r)|\psi_3\rangle +2 \chi^1_1\chi^1_4\langle\psi_1|v(r)|\psi_4\rangle\\
+2 \chi^1_2\chi^1_3\langle\psi_2|v(r)|\psi_3\rangle+2 \chi^1_2\chi^1_4\langle\psi_2|v(r)|\psi_4\rangle+2 \chi^1_3\chi^1_4\langle\psi_3|v(r)|\psi_4\rangle
\end{array}
\end{equation}
where the $\chi^i_j$'s are the projections of the truncated model space on
the basic vectors 2p, 3p1h. 
The procedure is re-iterated until the energy convergence has been obtained.

\section{Basic equations for the CFT of the (3p1h) CMWFs}
The normalization factors for the $\pi$ operators are calculated
by using the re-coupling algebra of Ref.~26). 
We obtain:
\begin{equation}\label{e32}
\begin{array}{l}
$Coef$_1
\displaystyle{
=\sum_{kJ'_rJ_r} \surd{[\hat{k}\hat{J}]} (-1)^{i+i'+k+J_2+J+1}
\left \{ \begin{array}{ccc}
J_r & k & J_1 \\
J & J_2 & J'_r \end{array}\right \}
(1-\delta_{j'2}\delta_{j1}(-1)^{j_1+j_2-J_i})\delta_{J_sJ_2},}
\end{array}
\end{equation}
\begin{equation}\label{e33}
\begin{array}{l}
$Coef$_2
\displaystyle{
=\sum_{kJ'_rJ^1_rJ^4_rJ_iJ_rJ_s} (\hat{J'_r}\hat{J_r^4})
\surd{[\hat{J_1}\hat{J_2}\hat{J_i}\hat{J^1_r}\hat{J_s}
\hat{J^4_r}\hat{k}\hat{J}]} 
(-1)^{i'+J'_r+J^1_r+J_r+J+1}}\\
\displaystyle{
\left \{ \begin{array}{ccc}
j_1 & j_2 & J_1 \\
J & J_2 & J'_r \end{array}\right \}
\left \{ \begin{array}{ccc}
j_3 & j_4 & J_2 \\
J'_r & j_1 & J^1_r \end{array}\right \}
\left \{ \begin{array}{ccc}
i & J^4_r & J'_r \\
J & j_2 & J_r \end{array}\right \}
\left \{ \begin{array}{ccc}
k & J_s & J^4_r \\
J_r & J & J \end{array}\right \}
\left \{ \begin{array}{ccc}
i & J'_r & J^4_r\\
i' & j_4 & J_s\\
J_i & J^1_r & k \end{array}\right \}}\\
\displaystyle{
  (1-\delta_{j'3}\delta_{j1}(-1)^{j_1+j_3-J_i}),   }
\end{array}
\end{equation}
and
\begin{equation}\label{e34}
\begin{array}{l}
$Coef$_3
\displaystyle{
=\sum_{kJ'_rJ^1_rJ^4_rJ_iJ_rJ_s}\surd{[\hat{J_i}\hat{J'_r}\hat{J_r}
\hat{J_s}\hat{k}\hat{J}]} 
(-1)^{i+i'+j_2+J'_r+J'_r}}\\
\displaystyle{
\left \{ \begin{array}{ccc}
j_1 & j_2 & J_1 \\
J & J_2 & J'_r \end{array}\right \}
\left \{ \begin{array}{ccc}
j_3 & j_4 & J_2 \\
J'_r & j_1 & J^1_r \end{array}\right \}
\left \{ \begin{array}{ccc}
i & J^4_r  & J'_r \\
J & j_1 & J_r \end{array}\right \}
\left \{ \begin{array}{ccc}
k & J_s & J^4_r \\
J_r & J & J \end{array}\right \}
\left \{ \begin{array}{ccc}
i & J'_r & J^4_r\\
i' & j_4 & J_s\\
J_i & J^1_r & k \end{array}\right \}}\\
\displaystyle{
 (1-\delta_{j'2}\delta_{j3}(-1)^{j_2+j_3-J_i}).  }
\end{array}
\end{equation}
By using these normalization coefficients we obtain the CFT for
the three partitions by diagonalizing the following matrix:  
\begin{equation}\label{e35}
\begin{array}{l}
\left ( \begin{array}{c|c|c}
$Coef$_1*$Coef$_1 & $Coef$_1*$Coef$_2 & $Coef$_1*$Coef$_3 \\ \hline
$Coef$_2*$Coef$_1 & $Coef$_2*$Coef$_2 & $Coef$_2*$Coef$_3 \\ \hline
$Coef$_3*$Coef$_1 & $Coef$_3*$Coef$_2 & $Coef$_3*$Coef$_3 
\end{array} \right )
\left ( \begin{array}{c}
|([a^{\dagger}_{j_1}a^{\dagger}_{j_2}]^{J_r}[a^{\dagger}_{j_3}a_{j_4}]^{J_s})^J\rangle\\
|([a^{\dagger}_{j_3}a^{\dagger}_{j_2}]^{J_r}[a^{\dagger}_{j_1}a_{j_4}]^{J_s})^J\rangle\\
|([a^{\dagger}_{j_1}a^{\dagger}_{j_3}]^{J_r}[a^{\dagger}_{j_2}a_{j_4}]^{J_s})^J\rangle
\end{array} \right )\\
=\left ( \begin{array}{c}
{^5}V^{\dagger(3,1)}_J(\alpha_2J_1J_2|\}\lambda_1\epsilon_1J_r\bar{\epsilon_1}J_s)\cdot
{^5}V^{(3,1)}_J(\alpha_2J_1J_2|\}\lambda_1 \epsilon_1J_r\bar{\epsilon_1}J_s)\\
{^5}V^{\dagger(3,1)}_J(\alpha_2J_1J_2|\}\lambda_2\epsilon_2J_r\bar{\epsilon_2}J_s)\cdot
{^5}V^{(3,1)}_J(\alpha_2J_1J_2|\}\lambda_2\epsilon_2J_r\bar{\epsilon_2}J_s)\\
{^5}V^{\dagger(3,1)}_J(\alpha_2J_1J_2|\}\lambda_3\epsilon_3J_r\bar{\epsilon_3}J_s)\cdot
{^5}V^{(3,1)}_J(\alpha_2J_1J_2|\}\lambda_3\epsilon_3J_r\bar{\epsilon_3}J_s)
\end{array} \right )
\left ( \begin{array}{c}
|([a^{\dagger}_{j_1}a^{\dagger}_{j_2}]^{J_r}[a^{\dagger}_{j_3}a_{j_4}]^{J_s})^J\rangle\\
|([a^{\dagger}_{j_3}a^{\dagger}_{j_2}]^{J_r}[a^{\dagger}_{j_1}a_{j_4}]^{J_s})^J\rangle\\
|([a^{\dagger}_{j_1}a^{\dagger}_{j_3}]^{J_r}[a^{\dagger}_{j_2}a_{j_4}]^{J_s})^J\rangle
\end{array} \right )
\end{array}
\end{equation}
The CFC coefficients associated to the $u$ operators are derived within the same
computational method introduced for the $\pi$ operators; the
normalization factors $(c_{\mu_i};\mu_i=4,5,6)$ are given below:
\begin{equation}\label{e36}
\begin{array}{l}
$Coef$_4
\displaystyle{
=\sum_{kJ'_rJ^1_rJ^2_rJ_iJ_rJ_s} \surd{[\hat{J_1}\hat{J_2}\hat{J^1_r}\hat{J_s}
\hat{J_i}\hat{k}\hat{J}]} 
(-1)^{i+j_1+j_3+J^2_r+J_2+J_r+J_s+J+1}}\\
\displaystyle{\left \{ \begin{array}{ccc}
j_1 & j_2 & J_1 \\
J'_r & j_4 & J^1_r \end{array}\right \}
\left \{ \begin{array}{ccc}
j_3 & j_4 & J_2 \\
J_1 & J & J'_r \end{array}\right \}
\left \{ \begin{array}{ccc}
i & j  & J_i \\
k & J^1_r & J^2_r \end{array}\right \}
\left \{ \begin{array}{ccc}
 i & J^2_r & J^1_r \\
J'_r & j_2 & J_s \end{array}\right \}
\left \{ \begin{array}{ccc}
J_s & J_r & J\\
J^2_r & j_4 & k\\
J'_r & j_3 & J \end{array}\right \}},
\end{array}
\end{equation}
\begin{equation}\label{e37}
\begin{array}{l}
$Coef$_5=$Coef$_4*(-1)^{j_1+j_2-J_1}~with~(j_1\to j_2),
\end{array}
\end{equation}
and
\begin{equation}\label{e38}
\begin{array}{l}
$Coef$_6
\displaystyle{
=\sum_{kJ'_rJ_r}\surd{[\hat{k}\hat{J}]}(-1)^{j_3+j_4+1+k+J_1+J_r}
\left \{ \begin{array}{ccc}
J_r & k & J_2 \\
J & J_1 & J'_r \end{array}\right \} \delta_{J_sJ_1}}
\end{array}
\end{equation}
By using these normalization coefficients we obtain the CFT for
the three partitions by diagonalizing the following matrix:  
\begin{equation}\label{e35a}
\begin{array}{l}
\left ( \begin{array}{c|c|c}
$Coef$_4*$Coef$_4 & $Coef$_4*$Coef$_5 & $Coef$_4*$Coef$_6 \\ \hline
$Coef$_5*$Coef$_4 & $Coef$_5*$Coef$_5 & $Coef$_5*$Coef$_6 \\ \hline
$Coef$_6*$Coef$_4 & $Coef$_6*$Coef$_5 & $Coef$_6*$Coef$_6 
\end{array} \right )
\left ( \begin{array}{c}
|([a^{\dagger}_{j_1}a_{j_4}]^{J_r}[a^{\dagger}_{j_2}a^{\dagger}_{j_3}]^{J_s})^J\rangle\\
|([a^{\dagger}_{j_2}a_{j_4}]^{J_r}[a^{\dagger}_{j_1}a^{\dagger}_{j_3}]^{J_s})^J\rangle\\
|([a^{\dagger}_{j_3}a_{j_4}]^{J_r}[a^{\dagger}_{j_1}a^{\dagger}_{j_2}]^{J_s})^J\rangle
\end{array} \right )\\
=\left ( \begin{array}{c}
{^5}U^{\dagger(3,1)}_J(\alpha_2J_1J_2|\}\mu_1\eta_1J_r\bar{\eta_1}J_s)\cdot
{^5}U^{(3,1)}_J(\alpha_2J_1J_2|\}\mu_1 \eta_1J_r\bar{\eta_1}J_s)\\
{^5}U^{\dagger(3,1)}_J(\alpha_2J_1J_2|\}\mu_2\eta_2J_r\bar{\eta_2}J_s)\cdot
{^5}U^{(3,1)}_J(\alpha_2J_1J_2|\}\mu_2\eta_2J_r\bar{\eta_2}J_s)\\
{^5}U^{\dagger(3,1)}_J(\alpha_2J_1J_2|\}\mu_3\eta_3J_r\bar{\eta_3}J_s)\cdot
{^5}U^{(3,1)}_J(\alpha_2J_1J_2|\}\mu_3\eta_3J_r\bar{\eta_3}J_s)
\end{array} \right )
\left ( \begin{array}{c}
|([a^{\dagger}_{j_1}a_{j_4}]^{J_r}[a^{\dagger}_{j_2}a^{\dagger}_{j_3}]^{J_s})^J\rangle\\
|([a^{\dagger}_{j_2}a_{j_4}]^{J_r}[a^{\dagger}_{j_1}a^{\dagger}_{j_3}]^{J_s})^J\rangle\\
|([a^{\dagger}_{j_3}a_{j_4}]^{J_r}[a^{\dagger}_{j_1}a^{\dagger}_{j_2}]^{J_s})^J\rangle
\end{array} \right )
\end{array}
\end{equation}
\section{A numerical application of the CFT}
In this appendix we apply the method of the previous appendix to 
calculate the CFC for a $(1d_{\frac{5}{2}}2s_{\frac{1}{2}})^2(d_{\frac{3}{2}}
p^{-1}_{\frac{3}{2}})^1$-(3p1h) CMWFs formed by coupling two particles assumed
to be in the $1d_{\frac{5}{2}}$ and $2s_{\frac{1}{2}}$ single particle shell model states
 to the $d_{\frac{3}{2}}p^{-1}_{\frac{3}{2}}$ p-h pair.
By using Eq.~(\ref{e32}) we write for 
$\mathrm{Coef}_1$:
\begin{equation}\label{e39}
\mathrm{Coef}_1
=-\sum_{J_r} 3.
\left \{ \begin{array}{ccc}
J_r & 1 & 2 \\
1 & 1 & 1 \end{array}\right \}
\end{equation}
 which by restricting the quantum numbers to k=1 for ($J_1=2,J_2=J_s=1,J=1,
J_r=2$) give the cases of Table~(III).
The $\mathrm{Coef}_2$ given in Eq.~(\ref{e33}) assumes for k=1 the form:
\begin{equation}\label{e42}
\begin{array}{l}
\displaystyle{\mathrm{Coef}_2}
\displaystyle{
=\sum_{J'_rJ^1_rJ^4_rJ_rJ_s}
(\hat{J'_r}\hat{J_r^4})[5\cdot3\cdot9\hat{J_i}\hat{J^1_r}\hat{J_s}]^{\frac{1}{2}} 
(-1)^{\frac{3}{2}+J'_r+J^1_r+J_r}}\\
\displaystyle{
\left \{ \begin{array}{ccc}
\frac{5}{2} & \frac{1}{2} & 2 \\
1 & 1 & J'_r \end{array}\right \}
\left \{ \begin{array}{ccc}
 \frac{3}{2} & \frac{3}{2} & 1 \\
J'_r & \frac{5}{2} & J^1_r \end{array}\right \}
\left \{ \begin{array}{ccc}
\frac{1}{2} & J^4_r & J'_r \\
1 & \frac{1}{2} & J_r \end{array}\right \}
\left \{ \begin{array}{ccc}
1 & J_s & J^4_r \\
J_r & 1 & 1 \end{array}\right \}
\left \{ \begin{array}{ccc}
\frac{1}{2} & J'_r & J^4_r\\
\frac{5}{2} & \frac{3}{2} & J_s\\
J_i & J^1_r & 1 \end{array}\right \}.}
\end{array}
\end{equation}
In Eq.~(\ref{e42}) the range of the indices in the sum is running over the 
following possibilities:
\begin{equation}\label{e43}
\begin{array}{|l|c|c|}\hline
 J'_r=3/2 &J'_r=5/2& J'_r=7/2\\
 J^1_r=0,1,2,3& J^1_r=1,2,3,4& J^1_r=2,3,4,5  \\
 J_s=0,1,2,3&  J_s=0,1,2,3& J_s=0,1,2,3\\
 J^4_r=1,2,3,4&  J^4_r=0,1,2,3,4,5 &  J^4_r=2,3,4,5\\
 J_r=0,1,2,3,4,5 &  J_r=0,1,2,3,4,5 & J_r=0,1,2,3,4,5  \\
 J_i=1,2,3,4&J_i=1,2,3,4& J_i=1,2,3,4\\ \hline
\end{array}
\end{equation}
By summing over all the possible cases we obtain for this special case the $\mathrm{Coef}_2$
given in Table~(IV).
The $\mathrm{Coef}_3$ coefficients are given in Eq.~(\ref{e34}) and for k=1 
we obtain: 
\begin{equation}\label{e44}
\begin{array}{l}
\displaystyle{\mathrm{Coef}_3}\\
\displaystyle{
=\sum_{J'_rJ^1_rJ^4_rJ_rJ_s}[\hat{J_i}\hat{J'_r}\hat{J_r}
\hat{J_s}3.3]^{\frac{1}{2}} 
(-1)^{\frac{1}{2}+J'_r+J^1_r}}\\
\displaystyle{
\left \{ \begin{array}{ccc}
\frac{5}{2} & \frac{1}{2} & 2 \\
 1 & 1 & J'_r \end{array}\right \}
\left \{ \begin{array}{ccc}
\frac{3}{2} & \frac{3}{2} & 1 \\
J'_r & \frac{5}{2} & J^1_r \end{array}\right \}
\left \{ \begin{array}{ccc}
 \frac{1}{2} & J^4_r  & J'_r \\
 J & \frac{5}{2} & J_r \end{array}\right \}
\left \{ \begin{array}{ccc}
1 & J^3_r & J^4_r \\
J_r & 1 & 1 \end{array}\right \}
\left \{ \begin{array}{ccc}
 \frac{1}{2} & J'_r & J^4_r\\
  \frac{3}{2}  & \frac{3}{2} & J_s\\
  J_i & J^1_r & 1 \end{array}\right \}.}
\end{array}
\end{equation}
For this special example the calculated $\mathrm{Coef}_3$ are given in Table~(III).
We see that the
matrix we have to diagonalize in order to get the cluster coefficients is
of the order of eight.
By introducing the coefficient of Table~(III,IV) in Eq.~(\ref{e35})
and by diagonalizing the derived matrix  we get the V-CFC given in Table~(V).
Analogous calculations can be performed for the u operators.
The $\mathrm{Coef}_4$ given in Eq.~(\ref{e36}) assumes for k=1 the form:
\begin{equation}\label{e45}
\begin{array}{l}
$Coef$_4
\displaystyle{
=\sum_{kJ'_rJ^1_rJ^2_rJ_iJ_rJ_s} [\hat{J_1}\hat{J_2}\hat{J^1_r}\hat{J_s}
\hat{J_i}\hat{k}\hat{J}]^{\frac{1}{2}} 
(-1)^{i+j_1+j_3+J^2_r+J_2+J_r+J_s+J+1}}\\
\displaystyle{\left \{ \begin{array}{ccc}
\frac{5}{2} & \frac{1}{2} & 2 \\
J'_r & \frac{3}{2} & J^1_r \end{array}\right \}
\left \{ \begin{array}{ccc}
\frac{3}{2} & \frac{3}{2} & 1 \\
 2  & 1 & J'_r \end{array}\right \}
\left \{ \begin{array}{ccc} 
 \frac{3}{2} & \frac{1}{2}   & J_i \\
  1 & J^1_r & J^2_r \end{array}\right \}
\left \{ \begin{array}{ccc}
 \frac{3}{2} & J^2_r & J^1_r \\
J'_r & \frac{1}{2} & J_s \end{array}\right \}
\left \{ \begin{array}{ccc}
J_s & J_s & 1\\
J^2_r & \frac{3}{2} & 1\\
J'_r & \frac{3}{2} & 1 \end{array}\right \}}.
\end{array}
\end{equation}
By summing over all possible partitions we obtain the $\mathrm{Coef}_4$
 coefficients which are given in Table~(\ref{a-VI}).
For the $\mathrm{Coef}_5$ we use the previous formula by replacing $j_1\to j_2$,
and we calculated coefficients given in Table~(VI).
The $\mathrm{Coef}_6$ are calculated form Eq.~(\ref{e38}) which 
in this special example takes the form:
\begin{equation}\label{e46}
\begin{array}{l}
$Coef$_6
\displaystyle{
=\sum_{J_r}3.*(-1)^{1+2+J_r}
\left \{ \begin{array}{ccc}
J_r & 1 & 1 \\
1 & 2 & 1 \end{array}\right \}.}
\end{array}
\end{equation}
Now since $J_r=1,2$ we have for $\mathrm{Coef}_6$ the coefficients given in Table
~(VII).
By using the coefficients $\mathrm{Coef}_i,i=4,5,6$
we derive the CFC for the U operators given in Table~(VIII).
By recalling that the pair coupled to $J_r$ is active
we write for the matrix elements calculated in the 
$((d_{\frac{5}{2}}s_{\frac{1}{2}})^2(d_{\frac{3}{2}}p^{-1}_{\frac{3}{2}})^1)^1$
CMWF the following value:
\begin{equation}\label{e47}
\begin{array}{l}
\langle d_{\frac{5}{2}}s_{\frac{1}{2}}d_{\frac{3}{2}}p^{-1}_{\frac{3}{2}}
|v(r)|d_{\frac{5}{2}}s_{\frac{1}{2}}d_{\frac{3}{2}}p^{-1}_{\frac{3}{2}}\rangle^{1}_a\\
=( 0.0617)^2\langle d_{\frac{5}{2}}s_{\frac{1}{2}}|v_{pp}(r)|d_{\frac{5}{2}}s_{\frac{1}{2}}\rangle^2_a
+( 0.0589)^2\langle d_{\frac{3}{2}}s_{\frac{1}{2}}|v_{pp}(r)|d_{\frac{3}{2}}s_{\frac{1}{2}}\rangle^2_a\\
+(-0.2088)^2\langle d_{\frac{3}{2}}s_{\frac{1}{2}}|v_{pp}(r)|d_{\frac{3}{2}}s_{\frac{1}{2}}\rangle^2_a
+( 0.2352)^2\langle d_{\frac{3}{2}}s_{\frac{1}{2}}|v_{pp}(r)|d_{\frac{3}{2}}s_{\frac{1}{2}}\rangle^2_a\\
+( 0.6029)^2\langle d_{\frac{5}{2}}d_{\frac{3}{2}}|v_{pp}(r)|d_{\frac{5}{2}}d_{\frac{3}{2}}\rangle^2_a
+(-0.3346)^2\langle d_{\frac{5}{2}}d_{\frac{3}{2}}|v_{pp}(r)|d_{\frac{5}{2}}d_{\frac{3}{2}}\rangle^2_a\\
+( 0.6466)^2\langle d_{\frac{5}{2}}d_{\frac{3}{2}}|v_{pp}(r)|d_{\frac{5}{2}}d_{\frac{3}{2}}\rangle^3_a
+( 0.0908)^2\langle s_{\frac{1}{2}}p_{\frac{3}{2}}|v_{ph}(r)|s_{\frac{1}{2}}p_{\frac{3}{2}}\rangle^2_a\\
+(-0.1504)^2\langle s_{\frac{1}{2}}p_{\frac{3}{2}}|v_{ph}(r)|s_{\frac{1}{2}}p_{\frac{3}{2}}\rangle^2_a
+( 0.4022)^2\langle d_{\frac{3}{2}}p_{\frac{3}{2}}|v_{ph}(r)|d_{\frac{3}{2}}p_{\frac{3}{2}}\rangle^1_a\\
+(-0.3278)^2\langle d_{\frac{3}{2}}p_{\frac{3}{2}}|v_{ph}(r)|d_{\frac{3}{2}}p_{\frac{3}{2}}\rangle^2_a
+( 0.7873)^2\langle d_{\frac{3}{2}}p_{\frac{3}{2}}|v_{ph}(r)|d_{\frac{3}{2}}p_{\frac{3}{2}}\rangle^2_a\\
+( 0.2094)^2\langle d_{\frac{5}{2}}p_{\frac{3}{2}}|v_{ph}(r)|d_{\frac{5}{2}}p_{\frac{3}{2}}\rangle^1_a
+( 0.1883)^2\langle d_{\frac{5}{2}}p_{\frac{3}{2}}|v_{ph}(r)|d_{\frac{5}{2}}p_{\frac{3}{2}}\rangle^2_a
\end{array}
\end{equation}
One has to note that in the ladder approximations
only the first seven terms of Eq.~(\ref{e47}) contribute to the matrix elements.
\section{Spectroscopic factors of the dressed wave functions}
The spectroscopic factor of the ground states of mass A=6 isotopes
is defined by:
\begin{equation}
S_{p_{\frac{3}{2}}p_{\frac{3}{2}}}=\langle\Phi_{2p}|a^{\dagger}_{p_{\frac{3}{2}}}a^{\dagger}_{p_{\frac{3}{2}}}|0\rangle
\end{equation}
The spectroscopic factor for the two neutrons in the $^6$He together with the
more significant components of the dressed J=$0^+$, T=1 wave function of the ground state
are given in Table IX.
For $^6$Li the spectroscopic factor and the more significant components
 of the $J=1^+$ T=0 ground state wave function are given in Table~X for 
 the small configuration space.
Corresponding values calculated with the large configuration space are given in Table~XI.
\section{Charge distribution for two correlated particles}
\begin{figure}
\includegraphics{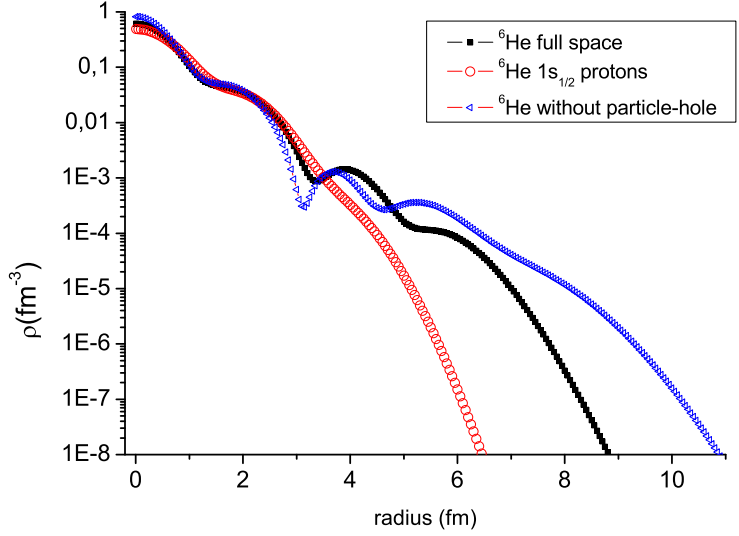}
  \caption{Calculated charge distributions of $^6$He: 
       black- calculated with a full configuration mixed base,
 blue- calculated without the particle-hole diagrams, and
       red- calculated for two protons in the $s_\frac{1}{2}$ shell.}
      \label{fig.1}
\end{figure}

\begin{figure}
\resizebox{0.55\textwidth}{!}
{\includegraphics{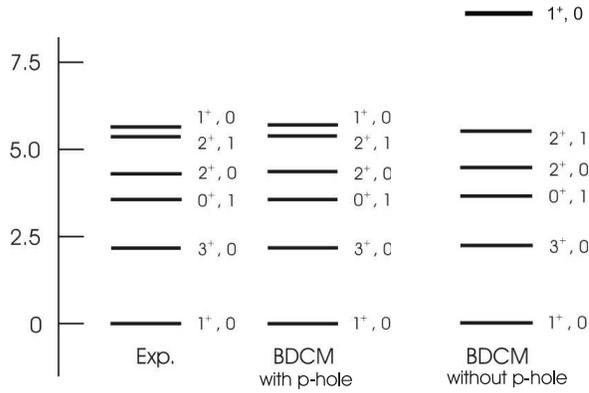}}
\vspace{2ex}
      \caption{Calculated spectrum of $^6$Li: 
      Left: energy levels calculated by allowing  
      the excitation of the 1s$_{\frac{1}{2}}$-
      and the 1p$_{\frac{3}{2}}$-hole.
      Right: energy levels
      calculated by restricting the hole to the 1s$_{\frac{1}{2}}$.}
      \label{fig.2}
      \end{figure}

\begin{figure}
\includegraphics{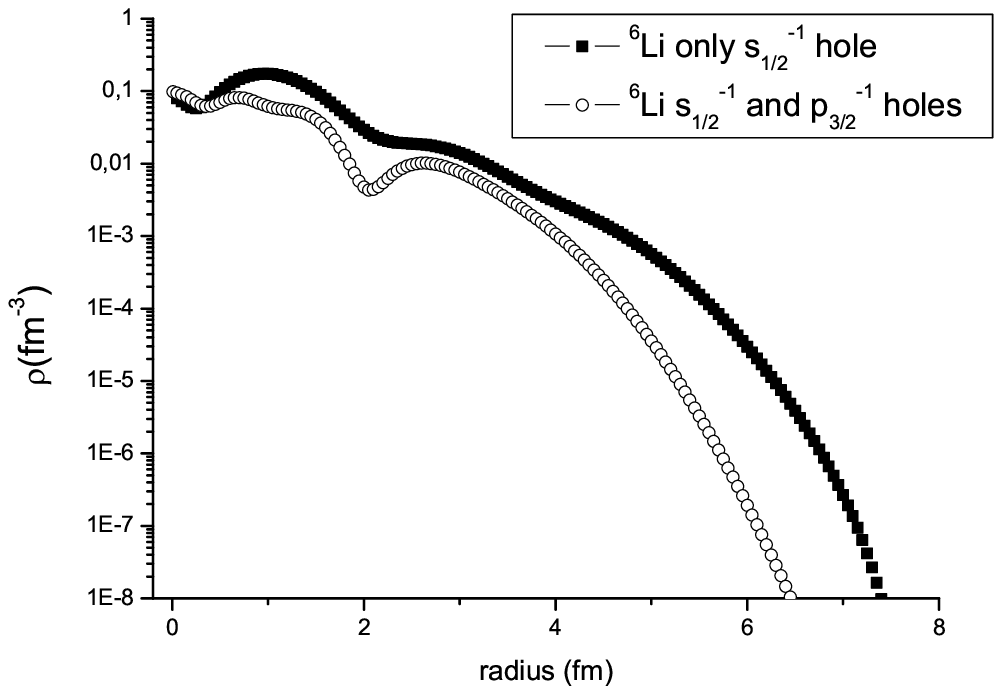}
\caption{Calculated charge distributions of $^6$Li: 
 black- calculated by restricting the hole to the $s_\frac{1}{2}$-state;
 white- calculated by considering $s_\frac{1}{2}$- and $p_\frac{1}{2}$-hole.
 The configuration space considered is formed
by 13 two-particle (shell model)-states and 512-(3p1h) states.}
\label{fig.3}
\end{figure}

In this appendix we give the difference between the distribution calculated
for two non-correlated particles in shell model states and that calculated 
for two dressed (correlated) particles.
In the shell model the distributions are evaluated by using the expectation values
of the operators :
\begin{equation}
\rho(r)=\sum_{\alpha}\langle\alpha|\delta(r-r_{\alpha})|\alpha\rangle a^{\dagger}_{\alpha}a_{\alpha}
\end{equation}
between the two particle shell model states:
\begin{equation}
[(j_1j_2)^J]=a^{\dagger}_{j_1}a^{\dagger}_{j_2}|0\rangle.
\end {equation}
By performing small algebra we obtain:
\begin{equation}
\rho(r)=\rho_{j_1}(r)+\rho_{j_2}(r),
\end{equation}
where $\rho_{j_i}$ is the single particle distribution.
This distribution is valid only in an extreme single particle model
i.e.: Shell Model, Hartree-Fock, mean field theories. 
In the BDCM the effect of the long range correlation must be included
consistently in the calculation of the densities.
We need therefore to calculate the distribution starting from the 
correlated particle pair given below:  
\begin{equation}\label{app1}
\tilde{\Psi}_{12}\equiv e^S\Psi_{12}=(1+S_1+S_2+S_3+\cdots)\Psi_{12},
\end{equation}
where the $S_i,i=1,2\cdots$ are the correlation operators of the i-th order.
Within our approximation the density should therefore be calculated from the model CMWFs: 
\begin{equation}
\tilde{\Psi}_{12}=
\sum_{ij}\chi_{ij}\Psi_{ij}+\sum_{ijkl}\chi_{ijkl}\Psi_{ij}\Psi_{kl}.  
\end{equation}
By using the amplitudes of Eq.~(\ref{e15a}) and the CFC of Eq.~(\ref{e16})
we derive, discarding the isospin quantum numbers, the following equation:
\begin{equation}\label{e48}
\begin{array}{l}
\displaystyle{
\rho^J(R)=(\sum_{\alpha_1}\chi^2_{\alpha_1}\sum_{nlNL\lambda}\langle n_1l_1n_2l_2\lambda|\}nlNL\lambda\rangle^2}
$angcoef$(\alpha_1\alpha_1\lambda\lambda J)\Phi^2_{NL}(R)\\
\displaystyle{
+2\sum_{\alpha_1\alpha_2}\chi_{\alpha_1} \chi_{\alpha_2}
(\sum_{J_r\epsilon_i}{^5}V^{(3,1)}_J(\alpha_2J_1J_2|\}\epsilon_iJ_r\bar{\epsilon}_iJ_s)}\\
\displaystyle{
\sum_{nlNL\lambda\lambda'} \langle n_1l_1n_2l_2\lambda|\}nlNL\lambda\rangle\langle n_il_in'_il'_i\lambda'|\}nlNL\lambda'\rangle}
  $angcoef$(\alpha_1\epsilon_i\lambda \lambda'J_r)\Phi^2_{NL}(R)\\
\displaystyle{
+\sum_{J_r\alpha_i}{^5}U^{(3,1)}_J(\alpha_2J_1J_2|\}\alpha_iJ_r\bar{\alpha}_iJ_s)}\\
\displaystyle{
\sum_{nlNL\lambda\lambda'}\langle n_1l_1n_2l_2\lambda|\}nlNL\lambda\rangle )\langle n_il_in'_il'_i\lambda'|\}nlNL\lambda'\rangle}
$angcoef$(\alpha_1\alpha_i\lambda \lambda'J_r)\Phi^2_{NL}(R))\\
\displaystyle{
+\sum_{\alpha_2\beta_2}\chi_{\alpha_2} \chi_{\beta_2}
(\sum_{J_rJ_{r'}\epsilon_i\epsilon_j}{^5}V^{\dagger(3,1)}_J(\alpha_2J_1J_2|\}\epsilon_iJ_r\bar{\epsilon}_iJ_s)
{^5}V^{(3,1)}_J(\beta_2J_1J_2|\}\epsilon_jJ'_r\bar{\epsilon}_jJ_s)}\\
\displaystyle{
\sum_{nlNL\lambda\lambda'}\langle n_il_in'_il'_i\lambda|\}nlNL\lambda\rangle\langle n_jl_jn'_jl'_j\lambda'|\}nlNL\lambda'\rangle}
$angcoef$(\epsilon_i\epsilon_j\lambda\lambda'J)\Phi^2_{NL}(R)\\
\displaystyle{
+\sum_{J_rJ_{r'}\beta_i\beta_j}{^5}U^{\dagger(3,1)}_J(\alpha_2J_1J_2|\}\beta_iJ_r\bar{\beta}_iJ_s)
{^5}U^{(3,1)}_J(\beta_2J_1J_2|\}\beta_jJ'_r\bar{\beta}_jJ_s)}\\
\displaystyle{
\sum_{nlNL\lambda\lambda'}\langle n_il_in'_il'_i\lambda|\}nlNL\lambda\rangle\langle n_jl_jn'_jl'_j\lambda'|\}nlNL\lambda'\rangle}
$angcoef$(\beta_i\beta_j\lambda\lambda'J)\Phi^2_{NL}(R))),
\end{array}
\end{equation}
where $R$ is the center of mass of the dressed particles, 
 (nlNL) the relative and center of mass angular momenta of the $(\alpha_i,\beta_i,\epsilon_i)$ pairs
, and the brackets are the Moshinski's brackets~\cite{mos01,urs01}. 
The $\chi_{\alpha_1}$ are the shell model amplitudes and the $\chi_{\alpha_2}$ the (3p1h)
amplitudes.
The angcoef's of Eq.~(\ref{e31}) are angular momentum transformation coefficients 
between the (jj) and the (ls) coupling given below:
\begin{equation}\label{e49}
\begin{array}{l}
$angcoef$(\alpha_1\alpha'_1\lambda\lambda'J)= (\frac{1}{2}\frac{1}{2}s,l_1l_2\lambda|\}(\frac{1}{2}l_1)j_1,
(\frac{1}{2}l_2)j_2,\lambda)(\frac{1}{2}\frac{1}{2}s',l'_1l'_2\lambda'|\}(\frac{1}{2}l'_1)j'_1,
(\frac{1}{2}l'_2)j'_2,\lambda')\\=
\surd{[\hat{j_1}\hat{j_2}]}\hat{\lambda}(-1)^{\lambda}
\left \{ \begin{array}{ccc}
\frac{1}{2}  &\frac{1}{2}  & s\\
l_1 & l_2 & \lambda\\
j_1 & j_2 & J_{12} \end{array}\right \}
\left \{ \begin{array}{ccc}
 L  & l  & \lambda\\
s & J_{12} & J \end{array}\right \}\\
\surd{[\hat{j'_1}\hat{j'_2}]}\hat{\lambda'}(-1)^{\lambda'}
\left \{ \begin{array}{ccc}
\frac{1}{2}  &\frac{1}{2}  & s'\\
l'_1 & l'_2 & \lambda'\\
j'_1 & j'_2 & J'_{1'2'} \end{array}\right \}
\left \{ \begin{array}{ccc}
 L'  & l'  & \lambda'\\
s' & J'_{1'2'} & J \end{array}\right \}.
\end{array}
\end{equation}
The angcoef$(\alpha_1\epsilon_1\lambda\lambda'J)$, angcoef$(\epsilon_1\epsilon'_1\lambda\lambda'J)$,
and  angcoef$(\beta_1\beta'_1\lambda\lambda'J)$ coefficients have a form analogous to that of Eq.~(\ref{e49}).
The symbols $\alpha_1$, $\alpha'_1$, $\beta_1,\beta_1'$, and $\epsilon_1,\epsilon_1'$ are given below:
\begin{equation}\label{50}    
\begin{array}{cc}
\alpha_1\longrightarrow&$two~valence~particles$\\
\alpha'_1, \beta_1, ~$and$~ \beta'_1 \longrightarrow&$two~particles~from~the~(3p1h)~CMWFs$\\
\epsilon_1 ~$and$~ \epsilon'_1\longrightarrow&$particle-hole~form~the~(3p1h)~CMWFs$
\end{array}
\end{equation}
The effect of the folded diagrams on the calculated distributions is given 
in Fig.~\ref{fig.1} where
we compare the correlated distribution of $^6$He calculated with
Eq.~(\ref{app1}) by neglecting the particle-hole diagrams with that
calculated by including also these diagrams. 
Analogous calculations will be performed for any operators.
This effect has been until now not considered by the other theoretical models.
The translational invariant density of Ref.~40)
is derived by assuming that the wave functions of the nuclei are
given in the terms of non correlated Slater's determinants.
           
\begin{table}[tbp]\label{b-I}
\begin{tabular}{lcccccccc}
\hline
hole          & $1s_{1/2}$ & & & & & & &   \\ 
energy        & -20.58     & & & & & & &   \\ \hline
hole/particle & $1p_{3/2}$ & & & & & & & \\ 
energy        & 1.43  & & & & & & &   \\ \hline
particle      &  $1p_{1/2}$ & $1d_{5/2}$ & $2s_{1/2}$ & $1d_{3/2}$
& $1f_{7/2}$  & $2p_{3/2}$ & $1f_{5/2}$ & $2p_{1/2}$ \\ \hline
energy        & 1.73       & 17.21 & 22.23 & 23.69
& 25.23       & 27.18      & 28.33 & 29.67 \\  \hline       
\end{tabular}
\caption{Single-particle scheme and single particle energies (MeV) used to form the model CMWFs}
\label{level-scheme}
\end{table}
\begin{table}[tbp]\label{b-II}
\label{Radius}
\begin{tabular}{lc}\hline
charge radius of $^6$He & Model \\ \hline
 1.944~\cite{nav03} & no-core shell model\\
 2.09~\cite{pip01}  & quantum Monte Carlo technique\\
 2.25~ this work             & $\mathrm{BDCM}$\\
 2.39~ this work&            $\mathrm{BDCM}$ without the folded diagrams\\
 2.06~this work              &$\mathrm{two~correlated}~1s_{\frac{1}{2}}-protons$\\
 1.99~\cite{wur97}     & Cluster\\   
 1.99~\cite{fun94}     & Cluster\\
 1.99~\cite{var94}     & Cluster\\
 2.054 $\pm$ .014\cite{wan04} & Isotopic Shift~(Exp.)\\  
\hline
\end{tabular}
\caption{Calculated charge radii for $^6$He in fm compared with the results 
obtained in other theoretical models and with the radius derived within the IS theory.}
\end{table}
\begin{table}[htp]\label{a-I}
\begin{tabular}{lcc}\hline
$J_r$ & $J_s$ & $\mathrm{Coef}_1$ \\ \hline
 2& 1& 0.0671 \\ \hline 
\end{tabular}
\caption{$\mathrm{Coef}_1$ as functions of $J_r$, $J_s$.} 
\end{table}
\begin{table}[htp]\label{a-II}
\begin{tabular}{lccc}\hline
 $J_r$& $J_s$ &  $\mathrm{Coef}_2$ & $\mathrm{Coef}_3$   \\  \hline  
     2&    1  & -0.5909001110$^{-01}$  & -0.1050328410$^{-02}$ \\
     2&    2  & -0.3650307710$^{-02}$  &  0.1547969910$^{-02}$ \\
     2&    3  & -0.2182041110$^{-01}$  &  -              \\
     3&    2  & -                & -0.1612873310$^{-02}$ \\ \hline     
\end{tabular}
\caption{$\mathrm{Coef}_2$ and $\mathrm{Coef}_3$ as functions of $J_r$, $J_s$.} 
\end{table}
\begin{table}\label{a-III}
\begin{tabular}{lcccc}\hline
 $\lambda_i$ & $J_r$ & $J_s$ & CFC                 & wave function\\ \hline
   1&2&1&0.0617 &$|((d_{\frac{5}{2}}s_{\frac{1}{2}})^{J_r=2}
(d_{\frac{3}{2}}p_{\frac{3}{2}}^{-1})^{J_s=1})^1\rangle$\\ 
   2&2&1& 0.0589 &$|((d_{\frac{3}{2}}s_{\frac{1}{2}})^{J_r=2}
(d_{\frac{5}{2}}p_{\frac{3}{2}}^{-1})^{J_s=1})^1\rangle$\\   
   3&2&2&-0.2088 &$|((d_{\frac{3}{2}}s_{\frac{1}{2}})^{J_r=2}
(d_{\frac{5}{2}}p_{\frac{3}{2}}^{-1})^{J_s=2})^1\rangle$\\  
   4&2&3& 0.2352 &$|((d_{\frac{3}{2}}s_{\frac{1}{2}})^{J_r=2}
(d_{\frac{5}{2}}p_{\frac{3}{2}}^{-1})^{J_s=3})^1\rangle$\\  
   5&2&1& 0.6029 &$|((d_{\frac{5}{2}}d_{\frac{3}{2}})^{J_r=2}
(s_{\frac{1}{2}}p_{\frac{3}{2}}^{-1})^{J_s=1})^1\rangle$\\ 
   6&2&2&-0.3346 &$|((d_{\frac{5}{2}}d_{\frac{3}{2}})^{J_r=2}
(s_{\frac{1}{2}}p_{\frac{3}{2}}^{-1})^{J_s=2})^1\rangle$\\
   7&3&2& 0.6466 &$|((d_{\frac{5}{2}}d_{\frac{3}{2}})^{J_r=3}
(s_{\frac{1}{2}}p_{\frac{3}{2}}^{-1})^{J_s=2})^1\rangle$\\  \hline     
\end{tabular}
\caption{Cluster factorization coefficients $V_1^{3,1}([d_{\frac{5}{2}}s_{\frac{1}{2}}J_1=2]
[d_{\frac{3}{2}}p_{\frac{3}{2}}^{-1}J_2=1]\}[j_ij_jJ_r][j_lj^{-1}_m J_s])$ 
calculated for the seven allowed partitions and related CMWFs.}
\end{table}
\begin{table}[htp]\label{a-IV}
\begin{tabular}{lccc} \hline
 $J_r$& $J_s$ & $\mathrm{Coef}_4$  & $\mathrm{Coef}_5$ \\   \hline 
     2&    1  &  0.3199068110$^{-01}$& 0.11957474   \\
     2&    2  &  0.5368016710$^{-01}$&   0.3506421710$^{-01}$  \\
     2&    3  &    -     &  -0.10351003 \\  \hline     
\end{tabular}
\caption{$\mathrm{Coef}_4$ and $\mathrm{Coef}_5$ as functions of $J_r$, $J_s$.} 
\end{table}
\begin{table}[htp]\label{a-V}
\begin{tabular}{lcc} \hline
$J_r$ & $J_s$ & $\mathrm{Coef}_6$ \\ \hline
1& 2& 0.5000 \\
2& 2& 0.0671  \\  \hline     
\end{tabular}
\caption{$\mathrm{Coef}_6$ as functions of $J_r$, $J_i$.}  
\end{table}
\begin{table}[htp]\label{a-VI}
\begin{tabular}{lcccc}\hline
 $\mu_i$ & $J_r$ & $J_s$ & CFC                 & wave function\\ \hline
   1&1&2& 0.0908 &$|((s_{\frac{1}{2}}p_{\frac{3}{2}}^{-1})^{J_r=2}
(d_{\frac{5}{2}}d_{\frac{3}{2}})^{J_s=1})^1\rangle$\\   
   2&2&2&-0.1574 &$|((s_{\frac{1}{2}}p_{\frac{3}{2}}^{-1})^{J_r=2}
(d_{\frac{5}{2}}d_{\frac{3}{2}})^{J_s=2})^1\rangle$\\  
   3&1&2& 0.4022 &$|((d_{\frac{3}{2}}p_{\frac{3}{2}}^{-1})^{J_r=1}
(d_{\frac{5}{2}}s_{\frac{1}{2}})^{J_s=2})^1\rangle$\\  
   4&2&2&-0.3278 &$|((d_{\frac{3}{2}}p_{\frac{3}{2}}^{-1})^{J_r=2}
(d_{\frac{5}{2}}s_{\frac{1}{2}})^{J_s=2})^1\rangle$\\ 
   5&2&3& 0,7873 &$|((d_{\frac{3}{2}}p_{\frac{3}{2}}^{-1})^{J_r=2}
(d_{\frac{5}{2}}s_{\frac{1}{2}})^{J_s=3})^1\rangle$\\ 
   6&1&2& 0.2094 &$|((d_{\frac{5}{2}}p_{\frac{3}{2}}^{-1})^{J_r=1}
(d_{\frac{3}{2}}s_{\frac{1}{2}})^{J_s=2})^1\rangle$\\
   7&2&2& 0.1883 &$|((d_{\frac{5}{2}}p_{\frac{3}{2}}^{-1})^{J_r=2}
(d_{\frac{3}{2}}s_{\frac{1}{2}})^{J_s=2})^1\rangle$\\ 
\hline
\end{tabular}
\caption{Cluster factorization coefficients 
$U_1^{3,1}([d_{\frac{5}{2}}s_{\frac{1}{2}}J_1=2]
[d_{\frac{3}{2}}p_{\frac{3}{2}}^{-1}J_2=1]\}[j_ij^{-1}_jJ_r][j_lj_mJ_s])$ 
calculated for the seven allowed partitions and related CMWFs.}
\end{table}
\begin{table}[htp]\label{a-VII}
\begin{tabular}{lcccc}\hline
$\mathrm{Spect.~fact.}$ &$1p_{\frac{3}{2}}1p_{\frac{3}{2}}$ & $1p_{\frac{1}{2}}1p_{\frac{1}{2}}$
 &$(1p_{\frac{3}{2}}1d_{\frac{5}{2}})^{2,1}(1p_{\frac{3}{2}}1s^{-1}_{\frac{1}{2}})^{2,0}$ 
 &$(1p_{\frac{3}{2}}1d_{\frac{5}{2}})^{2,1}(1p_{\frac{3}{2}}1s^{-1}_{\frac{1}{2}})^{2,1}$ \\ \hline
 0.9370 & 0.9680  & 0.1816 & -0.1145 & 0.0628 \\  \hline     
\end{tabular}
\caption{List of the most significant components of the ground state $0^+$,T=1
wave function of $^6$He with E=-24.97 MeV}
\end{table} 
\begin{table}[htp]\label{a-IX}
\begin{tabular}{lcccccc}\hline
$\mathrm{Spect.~fact.}$ &$1p_{\frac{3}{2}}1p_{\frac{3}{2}}$ & $1p_{\frac{3}{2}}1p_{\frac{1}{2}}$
  & $1p_{\frac{1}{2}}1p_{\frac{1}{2}}$ & $1d_{\frac{5}{2}}1d_{\frac{3}{2}}$  
 &$(1p_{\frac{3}{2}}1d_{\frac{5}{2}})^{2,0}(1p_{\frac{3}{2}}1s^{-1}_{\frac{1}{2}})^{1,0}$    
 &$(1p_{\frac{1}{2}}2s_{\frac{1}{2}})^{0,0}(1p_{\frac{3}{2}}1s^{-1}_{\frac{1}{2}})^{1,0}$ \\ \hline
 0.7649 & 0.8746  & 0.4559 & -.1009 & 0.0385 & 0.0261 & -0.0111 \\  \hline     
\end{tabular}
\caption{List of the most significant components of the ground state $1^+$,T=0
wave function of $^6$Li calculated within the small base with E=-19.30 MeV}
\end{table} 
\begin{table}[htp]\label{a-VIII}
\begin{tabular}{lcccccc}\hline
$\mathrm{Spect.~fact.}$ &$1p_{\frac{3}{2}}1p_{\frac{3}{2}}$ & $1p_{\frac{3}{2}}1p_{\frac{1}{2}}$
 &$1d_{\frac{5}{2}}1d_{\frac{5}{2}}$ &$1d_{\frac{5}{2}}1d_{\frac{3}{2}}$  
 &$(1p_{\frac{3}{2}}1d_{\frac{5}{2}})^{2,0}(1p_{\frac{3}{2}}1s^{-1}_{\frac{1}{2}})^{1,0}$    
 &$(1p_{\frac{1}{2}}2s_{\frac{1}{2}})^{0,0}(1p_{\frac{3}{2}}1s^{-1}_{\frac{1}{2}})^{1,0}$ \\ \hline
 0.6427 & 0.8017  & 0.4906 & -0.0636 & -0.1155 & 0.0653 & 0.03725 \\  \hline     
\end{tabular}
\caption{List of the most significant components of the ground state $1^+$,T=0
wave function of $^6$Li calculated within the large base with E=-21.2 MeV}
\end{table} 
\newpage
\newpage
\newpage

\begin{thebibliography}{99}
 \bibitem{vil63}
 F. Villars, {\it Proc. Enrico Fermi Int. School of Physics XXII}, Academic Press N.Y. (1961). 
 \bibitem{sha66} C.M. Shakin and Y.R. Waghmare, Phys. Rev. Lett. 16, 403 (1966);
C.M. Shakin, Y.R. Waghmare, and M.H. Hull,
Phys. Rev. 161, 1006 (1967).
 \bibitem{tom01} M. Tomaselli, L.C. Liu, S. Fritzsche, T. K{\"u}hl,
and D. Ursescu, Nucl. Phys. A738, 216 (2004);
 M. Tomaselli, L.C. Liu, S. Fritzsche, T. K{\"u}hl,
J. Phys. G: Nucl. Part. Phys. 30, 999 (2004).
 \bibitem{bru01}
K.A. Br{\"u}ckner, {\it The Many Body Problem}, John Wiley and Sons, New York, 1959.
 \bibitem{kuo01}
T.T.S. Kuo and E. Osnes, {\it Folded-Diagrams Theory of the Effective Interaction
in Atomic Nuclei}, Springer Lecture Notes in Physics vol. 366, Berlin (1991).
 \bibitem{nav98} P. Navr$\acute{\mathrm{a}}$til and B.R. Barrett, Phys. Rev. C57, 3119 (1998);
 P. Navr$\acute{\mathrm{a}}$til and E. Caurier, Phys. Rev. C69, 014331 (2004).
 \bibitem{sta05}
I. Stancu, B.R. Barrett, P. Navr$\acute{\mathrm{a}}$til, and J.P. Vary, 
Phys. Rev. C71, 044325 (2005).
 \bibitem{Kan97} Y. KanadaEn'yo and Horiuchi, Phys. Rev. C55, 2860 (1997).
 \bibitem{pip01} S.C. Pieper and R.B. Wiringa, Ann. Rev. Part. Sci. 51, 53 (2001). 
 \bibitem{wan04} L.-B. Wang, P. M{\"u}ller, V. Bailey {et al.}, 
 Phys. Rev. Lett. 93, 142501 (2004).
 \bibitem{wur97}
J. W{\"u}rzer and H.M. Hofmann, Phys. Rev. C55, 688, (1997).
 \bibitem{fun94}
S. Funada, H. Kaneyama and Y. Sakuragi, Nucl. Phys. A575, 93 (1994).
 \bibitem{var94}
K. Varga, Y. Suzuki and Y. Ohbayasi, Phys. Rev. C50, 189 (1994).
\bibitem{tom02}
M. Tomaselli, T. K{\"u}hl, W. N{\"o}rtersh{\"a}user, G. Ewald, R. Sanchez, S. Fritzsche,
and G.S. Karshenboim, Can. J. Phys. 80, 1347 (2002). 
\bibitem{ewa05}
 G. Ewald, W. N{\"o}rtersh{\"a}user, A. Dax,  {\it et al.}, Phys. Rev. Lett. 93, 113002 (2004);
B.A. Bushaw, W. N{\"o}rtersh{\"a}user, G. Ewald, {\it et al.}, Phys. Rev. Lett. 91, 043004 (2003); 
R. S{\'a}nchez, W. N{\"o}rtersh{\"a}user, G. Ewald, D. Albers, J. Behr, P. Bricault, B. A. Bushaw, 
A. Dax, J. Dilling, M. Dombsky, G. W. F. Drake, S. G{\"o}tte, R. Kirchner, 
H.-J. Kluge, T. K{\"u}hl, J. Lassen, C.D.P. Levy, M.R. Pearson, E.J. Prime, 
V. Ryjkov, A. Wojtaszek, Z.-C. Yan, C. Zimmerman, Phys. Rev. Lett. 96, 033002 (2005).
 \bibitem{dra00}
 Z.-C. Yan and G.W.F. Drake, Phys. Rev. A66, 042504 (2002).
 \bibitem{li071}
G.C. Li, I. Sick, R.R. Whitney, and M.R. Yearian, Nucl. Phys. A162, 583 (1971).
 \bibitem{auf82}
 P. Aufmuth, J. Phys. B: At. Mol. Phys., 15, 3127 (1982).
 \bibitem{tom05} 
 M. Tomaselli, Can. J. of Phys. 83, 467 (2005). 
\bibitem{mar01}
 M. Anguiano and G. Co, J. Phys. G27, 2109 (2001).
\bibitem{lal96}
 G.A. Lalazissis and S. E. Massen, arXiv:nucl/9601014vl (1996).
\bibitem{str80}
 O. Bohigas and S. Stringari, Phys. Lett. B95, 9 (1980).
\bibitem{ari02}
H. Noya, A.Arima, and H. Horie, Prog. Part. Nucl. Phys. 1, 41 (1958). 
 \bibitem{mil75} 
D.J. Millener and D. Kurath, Nucl. Phys. 255, 315 (1975).
 \bibitem{bro64} 
G.E. Brown, {\it Unified Theory of Nuclear Models},
Amsterdam: North-Holland (1964). 
 \bibitem{rac61} 
U. Fano and G. Racah, {\it Irreducible Tensorial Sets}, Academic Press New York (1959); 
G. Racah, {\it Group theory and spectroscopy},
CERN Report 61-8,  Geneve, Schwitzerland (1961).
 \bibitem{til01}
D.R. Tilley, C. M. Cheves, J.L. Gowdin, G. M. Hale {\it et al.}, Nucl. Phys. A708, 3 (2002);
 D.R. Tilley, J.H. Kelley, J.L. Godwin, D.J. Millener, J.E. Purcell, 
{\it et al.}, Nucl. Phys. A745, 155 (2004); H. Koure and M. Yamada, Nucl. Phys. A671, 96(2000);
N. Walet, ``Nuclear and Particle Physics'',
http://walet.Phy.umist.ac.uk/P615/Notes.pdf, pg. 38 (2003). 
 \bibitem{zhe94}
 D.C. Zheng, J.P.Vary ,and B.R. Barrett, Phys. Rev. C50, 2841 (1994).
 \bibitem{sha67} C.M. Shakin, Y.R. Waghmare, M. Tomaselli, and, M.H. Hull,
Phys. Rev. 161, 1015 (1967).
\bibitem{kuo02}
 T.T.S. Kuo, H. M{\"u}ther, and K. Amir Azimi-Nili, 
Nucl. Phys. A606, 15 (1996).
 \bibitem{jia01}
M.F. Jiang, R. Machleidt, D.B. Stout, and T.T.S. Kuo, Phys. Rev. C46, 910 (1992);
M. Lacombe, B. Loiseau J.M. Richard, R. Vinh Mau, 
J. C${\hat{\mathrm o}}$t${\acute{\mathrm e}}$, 
P. Pir${\grave{\mathrm e}}$s, and R. de Tourreil, Phys. Rev. C21, 861 (1980);
R. Machleidt, Adv. Nucl. Phys. 19, 189 (1989);
S. Bogner, T.T.S. Kuo, L. Coraggio, A. Covello, and N. Itaco, 
Phys. Rev. C65, 051301(R) (2002).  
 \bibitem{bar61}
E. Baranger and C.W. Lee, Nucl. Phys. 22, 157 (1961).
 \bibitem{unn58}
I. Unna and I. Talmi, Phys. Rev. 112, 452 (1958).
 \bibitem{nav03} 
P. Navr$\acute{\mathrm{a}}$til and E.W. Ormand, Phys. Rev C68, 034305 (2003).
 \bibitem{aks02}
F. Aksouh, PHD-Thesys, IPNO-T-03-02. Paris-sud University, Orsay (2002). 
 \bibitem{ajz84} 
F. Ajzenberg-Selove, Nucl. Phys. A490, 1 (1988). 
 \bibitem{bro01}
B.A. Brown, Prog. Part. Nucl. Phys. 47, 524 (2001). 
 \bibitem{mos01}
 T.A. Brody and M. Moshinski, ``Table of Transformation Brackets'', 
Monografias del Instituto de Fisica, Universidad Nacional Autonoma de Mexico (1960). 
 \bibitem{urs01}
 D. Ursescu, M. Tomaselli, T. K{\"u}hl, and S. Fritzsche, Comput. Phys. Comm. 173, 140 (2005).
\bibitem{nav04}
P. Navr$\acute{\mathrm{a}}$til, Phys. Rev. C70, 014317 (2004).
\end{thebibliography}
             \end{document}